\documentclass[a4paper,fleqn,colorlinks=true, allcolors=blue]{cas-dc}

\usepackage[numbers,sort&compress]{natbib}
\usepackage{url}
\usepackage{hyperref}
\usepackage{xurl}   

\def\tsc#1{\csdef{#1}{\textsc{\lowercase{#1}}\xspace}}
\tsc{WGM}
\tsc{QE}
\tsc{EP}
\tsc{PMS}
\tsc{BEC}
\tsc{DE}
\usepackage[T1]{fontenc}
\usepackage{placeins}
\usepackage{lmodern}
\usepackage{tabularx}
\usepackage{array}
\usepackage{ragged2e}
\newcolumntype{Y}{>{\RaggedRight\arraybackslash}X}
\usepackage{comment}
\usepackage{microtype}
\usepackage[utf8]{inputenc}
\usepackage{pifont}
\newcommand{\cmark}{\text{\ding{51}}}     
\newcommand{\xmark}{\text{\ding{55}}}     
\newcommand{\tmarksym}{\ding{115}}
\usepackage[table]{xcolor}
\usepackage{geometry}
\usepackage{caption}
\usepackage{amsmath}
\usepackage{graphicx}
\usepackage{rotating}

\begin{document}
\let\WriteBookmarks\relax
\def\floatpagepagefraction{1}
\def\textpagefraction{.001}
\shortauthors{M. Zayene et~al.}

\title[mode = title]{Toward hyper-adaptive AI-enabled 6G networks for energy efficiency: techniques, classifications and tradeoffs}

\author[1]{Mariem Zayene}[type=editor,
                        auid=000,bioid=1,
                        orcid=0000-0001-0000-0000]
\cormark[1]
\ead{mariem.zayene@uca.fr}

\credit{Conceptualization of this study, Methodology, Software}

\affiliation[1]{organization={LIMOS, Université de Clermont-Auvergne},
                addressline={}, 
                postcode={63178}, 
                city={Clermont-Ferrand},
                country={France}}

\author[1]{Oussama Habachi}

\author[1]{Gérard Chalhoub}

\cortext[cor1]{Corresponding author}
\cortext[cor2]{Principal corresponding author}

\begin{abstract}
Energy efficiency is shaping up to be one of the most challenging issues for 6G networks. The reason is fairly straightforward: Networks will need to meet extreme service demands while remaining sustainable and traditional optimization techniques are too limited. With users moving, traffic swinging unpredictably and services pulling in different directions, management has to be adaptive and AI may offer a way forward. This survey looks at how well AI-based methods actually deliver on that promise. We organize the review around practical use cases. For each use case, we examine how AI techniques contribute to feedback-driven adaptability and rapid decision-making under dynamic conditions. We then evaluate them against seven central dynamic aspects that we consider unavoidable in 6G. The survey also discusses crucial tradeoffs between energy efficiency and the remaining 6G main objectives such as latency, reliability, fairness and coverage, and finally identifies gaps and future research directions. 
\end{abstract}


\maketitle

\section{Introduction}
The future 6G network foresees a radical shift in world connectivity through the unified integration of disparate networks. This unification combines terrestrial, aerial, satellite and underwater telecommunication systems with prospecting technologies such as IoT, edge computing and integrated sensing \cite{surveyB}. Through ubiquitous and incessant connectivity, 6G plans to overcome conventional limitations between networks providing unified services across a wide range of scenarios from highly populated urban regions to remote and hostile areas. This unification not only provides resilient connectivity but also facilitates multi-faceted applications such as real-time remote monitoring, autonomous systems and immersed multimedia experience. This paves the way for 6G at the foundation of the next technological revolution.

Nevertheless, the growing heterogeneity of 6G use-cases and architectural components imposes not only never-before-seen complexity but also high dynamicity on communication environments. The explosion of mobile devices, IoT usage and integrated technologies requires adaptive and intelligent solutions to address such challenges efficiently. As the relative stability of earlier infrastructure-centric communication systems gives way to dynamic 3D-covering communications, model-based approaches become gradually inadequate for dealing with complexities of such environments. These approaches fail miserably in adaptability to fast changes and uncertainty which are inherently anchored within 6G systems.

Artificial intelligence (AI)/ Machine learning (ML) provides a fundamental answer for such dynamic challenges. In contrast to established techniques, ML data-driven capabilities help inferring what the network conditions are, predicting user actions, and optimizing resource usage on the fly. By drawing deeper insight from enormous volumes of disparate datasets, ML is capable of dynamically controlling network traffic, service needs, and channel changes, facilitating intelligent traffic control, automatic redesign, and enhanced protection functions. This flexibility is crucial to sustain optimum performance, typified here by extreme mobility, very dense deployment of devices, and varying service needs respectively. As 6G extends the reach of connectivity and integration, ML is destined to ensure that the network should not only be intelligent, but resilient and efficient, despite unprecedented dynamism.

There are two main concepts defining the incorporation of AI into communication systems: AI for network and network for AI \cite{pan2021network}. Indeed, AI for network is about applying AI technologies to improve and automate many aspects of network management such as optimization of the network, security strengthening and resource utilization management. Network for AI, on the other hand, deals with designing, optimizing and building the network infrastructure for the support of AI workloads. These workloads tend to require high data movement with high-latency avoidance. Possible considerations include high-bandwidth links, low-latency communication, scalability and integration with edge computing, bringing AI processing closer to where data is generated to cut down on delay and bandwidth use.

In this paper, we focus on the use of AI for networks to address emerging challenges and improvements particularly in the energy efficiency of 6G systems.

\subsection{From 1G to 6G: A progressive path toward network adaptability}

The progression of cellular networks from 1G through 6G is a linear transformation of the design philosophy from fixed hardware-bound infrastructures to smart adaptive ecosystems. In fact, early generations 1G through 3G had limited flexibility with fixed protocols and circuit-switched architectures mainly for voice or simple data services. The network performance was mostly pre-specified and fixed at a manual level without context awareness or dynamic performance optimization.

With 4G, the movement towards packet-switching networks and software-defined components implemented crude adaptations at their onset. Although using dynamic scheduling and QoS differentiation, it was still limited to respond efficiently to traffic changes. Indeed, the adaptation remained rule-based and centrally managed lacking both predictive capability and autonomy.

5G advanced one step further with the addition of network slicing, ultra-reliable low-latency communications (URLLC), and support for various service profiles such as enhanced Mobile Broadband (eMBB) and massive Machine Type Communications (mMTC). Although such features target flexibility, the mechanisms were semi-static: Slices are generally pre-configured and  responsiveness towards dynamic changes (e.g., mobility, energy states or security threats) is not deeply integrated into the architecture. Furthermore, AI is seen as an auxiliary enhancement on top, not a part of the architecture foundation.

On the other hand, 6G is visualized as the first-ever adopted hyper-adaptivity-centric design approach. Adapting is no longer a choice but a necessity due to the requirements for extreme heterogeneity, dynamic topologies, energy limitations and dynamic adversarial attacks. Through native AI convergence, edge intelligence, software-defined networking and integrated sensing, 6G networks target autonomous operation aware of the context. The 6G networks will continue to adapt dynamically to real-time fluctuations either in user activity, networking or service requirements. This intelligent adaptability is destined to be the signature of wireless communicators of the future.

\subsection{Visionary architectures for adaptive 6G networks}

In addition to academic research, a number of industry and standardization organizations have issued foresighted visions for 6G that center adaptability on next generation network design. These visionary contexts do not illustrate specific technical deployments but provide architectural visions and strategic agendas that significantly highlight intelligent, real-time and context-aware functions as being strongly needed. Specifically, they emphasize adaptability as a way of enabling next-generation solutions to cope with the dynamic and complex nature of future wireless environments such as heterogeneous infrastructure, mobility-induced variations, changing service needs and new security threats.

As our survey aims at assessing the energy efficiency adaptability of AI-based 6G, for the rest of this sub-section, we study some sample 6G vision papers from Huawei, Nokia (Hexa-X) and Ericsson in order to explore how the industry plans to design adaptability towards a sustainable 6G network.


\paragraph{Nokia – Hexa-X 6G Vision.}
Nokia, under its vision-lead for the Hexa-X and Hexa-X-II flagship European projects, puts sustainability and adaptability at the heart of its 6G vision. The Hexa-X key findings make sustainability a top-level Key Value Indicator (KVI) along with performance and trustworthiness, with clear energy efficiency targets quantified in Joule-per-bit~\cite{Uusitalo2021HexaXKeyFindings,HexaX-D1.4}. Nokia's 6G system architecture proposals outline AI-native, modular and adaptable design in a way that network functions, protocol layers, or even physical components could be dynamically adapted to service needs, device limitations and environmental factors~\cite{Nokia2020TechInnovations,Nokia2025SystemArchitecture}. Software-defined functions, omnipresent machine learning and distributed control enable closed-loop optimization across network abstraction layers. In terms of energy efficiency, Hexa-X makes a prime plea for "energy efficiency by design" based on AI-aiding workload orchestration, energy-sensitive radio design, flexible resource management to reduce consumption across a wide range of scenarios~\cite{HexaX-D1.4,HexaXII-D2.6}. Together, these concepts emphasize Nokia’s perspective that AI-driven adaptability is crucial towards intelligent, scalable, and sustainable 6G networking.

\paragraph{Ericsson – 6G RAN Energy Vision.}

Ericsson’s 6G vision especially in its 2024 white paper on Energy performance of 6G Radio Access Networks: a once-in-a-decade opportunity focuses on creating an adaptive energy-aware RAN scaling activities to meet demand with less wasteful baseline power consumption~\cite{Ericsson2024RANEnergy}. The paper stresses that a large majority of components in today’s cellular systems draw almost fixed power even under light load that 6G needs to break this wastefulness by redesigning coverage overhead, mandatory transmissions and idle-mode signaling. In follow-on work, Ericsson highlights that energy performance on both the network side and the device side is a number-one 6G priority with a suggestion that 6G should have quantitative energy consumption specifications relative to preceding generations~\cite{Ericsson2024ThreeDimRequirements}. As one of the architectural recommendations, there is a suggestion of separating uplink carrier selection from downlink carrier selection for a better energy-optimal coverage-throughput trade-off~\cite{Ericsson2025StandardizationStep}. Overall, Ericsson sets out energy efficiency as a holistic system goal, pushing for permanent measurement, dynamic configuration, and intelligence at all levels for the attainment of a more sustainable 6G RAN~\cite{Ericsson2025AISustainability}.

\paragraph{Huawei – "6G: The Next Horizon" Vision (2023).}
The Next Horizon document on Huawei’s 6G vision outlines a whole-scene architecture founded on six pillars that are centered on native AI, sustainability, and trustworthiness~\cite{Huawei2023}. Under this vision, communications, sensing, computing and AI profoundly merge into a smart cyber-physical ecosystem. Adaptability is integrated into this ecosystem: The vision of Huawei is AI capabilities distributed across network layers for enabling real-time system awareness, self-optimization and context-aware orchestration.

On the energy side, Huawei aims at a 100× overall energy efficiency improvement from 5G, primarily through smart energy control and dynamic resource scaling instead of solely rely on hardware upgrades~\cite{Huawei2023,NativeTrustworthiness}. Their vision encompasses a hybrid ground, satellite and air node network enabling dynamic reconfiguration of connect paths for routing traffic in less energy-consuming ways. They also encourage service-based slicing and energy-perceptive orchestration with resource scheduling driven by user needs, tolerable latency, and device limitations.

Huawei’s perspective promotes adaptability from a performance booster to a core architectural requirement: for a future 6G network, energy management needs to be carried out under a context-aware, self-optimizing and resilient regime. These suggestions highlight that adaptability driven by AI is not a desirable addition but fundamental to enabling sustainable and intelligent 6G infrastructures.

\subsection{Why an adaptability focused AI survey is needed?}
Recent 6G visions put forward by major industry players such as Huawei, Nokia, and Ericsson leave little doubt that adaptability will form a cornerstone of next-generation network design. However, these visions remain largely at the architectural or conceptual level. They underline the importance of intelligent, energy-efficient and secure-by-design infrastructures capable of real-time reconfiguration but rarely venture into the specific AI/ML mechanisms that might make such adaptability possible.

At the same time, academic research has generated a wide range of surveys on AI and ML for wireless networks. Still, as discussed in the following section, much of this literature either revisits conventional AI applications or treats emerging 6G scenarios in a rather static and fragmented way. What is often missing is a direct question into a key issue: how far can AI-driven methods truly adapt to the dynamic multidimensional nature of 6G environments?

This survey takes this question as its starting point. It offers an adaptability-focused examination of AI and ML techniques within one of the most demanding 6G domains: energy efficiency. By relating the reviewed solutions to both the evolving operational realities of 6G and the broader strategic visions of industry and standardization bodies, we seek to clarify how AI can move beyond a supporting role to become a fundamental enabler of truly adaptive 6G systems.

The next section reviews existing survey literature to assess how these challenges have been addressed so far and to identify the remaining gaps.

\section{Related Work}

\begin{table*}[htbp]
\rotatebox{90}{%
\begin{minipage}{\textheight}  
\centering
\caption{Comparison of existing surveys with respect to 8 criteria. Symbols: \cmark = fully addressed, \tmarksym = partially addressed, \xmark = not addressed.}
\label{tab:comparative-surveys}
\renewcommand{\arraystretch}{1.2}
\begin{tabular}{lcccccccc}
\hline
\textbf{Survey} & \textbf{6G focus} & \textbf{EE focus} & \textbf{Use-case org.} & \textbf{AI by goal} & \textbf{Adapt. to dynamics} & \textbf{Explicit dynamics} & \textbf{Adapt. mapping} & \textbf{Tradeoff analysis} \\
\hline
\multicolumn{9}{l}{\textbf{Dedicated AI-for-EE surveys}} \\
\cite{survey5}   & \cmark & \cmark & \xmark & \tmarksym & \xmark & \xmark & \xmark & \tmarksym \\
\cite{surveyN5}  & \tmarksym & \cmark & \xmark & \xmark & \xmark & \xmark & \xmark & \tmarksym \\
\cite{surveyN6}  & \cmark & \cmark & \xmark & \tmarksym & \xmark & \xmark & \xmark & \tmarksym \\
\cite{surveyN8}  & \cmark & \cmark & \tmarksym & \xmark & \tmarksym & \tmarksym & \xmark & \tmarksym \\
\hline
\multicolumn{9}{l}{\textbf{Vision and framework-oriented surveys}} \\
\cite{survey12}  & \cmark & \tmarksym & \xmark & \tmarksym & \xmark & \xmark & \xmark & \xmark \\
\cite{survey13}  & \cmark & \tmarksym & \xmark & \tmarksym & \xmark & \xmark & \xmark & \xmark \\
\cite{surveyN1}  & \cmark & \cmark & \xmark & \xmark & \xmark & \xmark & \xmark & \xmark \\
\cite{surveyN3}  & \cmark & \tmarksym & \xmark & \xmark & \xmark & \xmark & \xmark & \xmark \\
\cite{surveyB}   & \cmark & \tmarksym & \xmark & \xmark & \xmark & \xmark & \xmark & \xmark \\
\hline
\multicolumn{9}{l}{\textbf{Domain-specific surveys (partial EE focus)}} \\
\cite{survey6}   & \cmark & \cmark & \tmarksym & \xmark & \xmark & \xmark & \xmark & \xmark \\
\cite{survey2}   & \tmarksym & \cmark & \tmarksym & \xmark & \xmark & \xmark & \xmark & \xmark \\
\cite{survey11}  & \cmark & \cmark & \tmarksym & \tmarksym & \xmark & \xmark & \xmark & \xmark \\
\cite{survey7}   & \cmark & \tmarksym & \tmarksym & \xmark & \xmark & \xmark & \xmark & \xmark \\
\cite{survey8}   & \cmark & \cmark & \tmarksym & \xmark & \xmark & \xmark & \xmark & \xmark \\
\cite{survey9}   & \cmark & \cmark & \tmarksym & \tmarksym & \tmarksym & \tmarksym & \xmark & \tmarksym \\
\cite{surveyN7}  & \cmark & \cmark & \tmarksym & \xmark & \tmarksym & \tmarksym & \xmark & \cmark \\
\cite{surveyG}   & \tmarksym & \tmarksym & \tmarksym & \xmark & \xmark & \xmark & \xmark & \xmark \\
\hline
\multicolumn{9}{l}{\textbf{General 6G surveys with broad scope}} \\
\cite{survey1}   & \cmark & \tmarksym & \xmark & \tmarksym & \xmark & \xmark & \xmark & \xmark \\
\cite{survey3}   & \cmark & \xmark & \xmark & \tmarksym & \xmark & \xmark & \xmark & \xmark \\
\cite{survey10}  & \cmark & \tmarksym & \xmark & \xmark & \xmark & \xmark & \xmark & \xmark \\
\cite{surveyA}   & \cmark & \tmarksym & \tmarksym & \xmark & \xmark & \xmark & \xmark & \xmark \\
\cite{surveyD}   & \tmarksym & \xmark & \xmark & \tmarksym & \xmark & \xmark & \xmark & \xmark \\
\cite{surveyE}   & \tmarksym & \xmark & \xmark & \tmarksym & \xmark & \xmark & \xmark & \xmark \\
\cite{surveyF}   & \tmarksym & \xmark & \xmark & \xmark & \xmark & \xmark & \xmark & \xmark \\
\hline
\textbf{Our survey} & \cmark & \cmark & \cmark & \cmark & \cmark & \cmark & \cmark & \cmark \\
\hline
\end{tabular}

\end{minipage}
}
\end{table*}
Several recent survey articles have addressed the interplay between AI and 6G networks with different focus on energy efficiency (EE), conceptual vision and enabler technologies. We categorize these works into four groups to summarize their contributions and limitations with respect to our current survey.

\subsection{Dedicated AI-for-EE Surveys}
This category includes surveys that explicitly target energy efficiency in 6G networks through the use of AI and ML techniques. Unlike broader works that only mention EE in passing, these studies place EE at the center of their analysis and investigate how AI methods can reduce power consumption at different layers of the network. They differ in scope and depth, ranging from general overviews of AI-based EE strategies to highly focused analyses of specific paradigms such as federated learning (FL) or distributed ML.

A technical overview is provided in \cite{survey5} which is a specific directed work that addresses machine learning techniques on energy efficiency in base stations and access networks. It surveys the supervised, the unsupervised and the reinforcement learning models and describes how they can decrease the power consumption through dynamic resource allocation, traffic forecasting and load balancing. It presents examples on ML-based EE solutions on the level of the radio access and the core network making this work a useful guide towards practical EE approaches. However, no framework is given on how effectively they will adapt under the fast-changing environments that will be the reality under the realities of the 6G.

Some surveys narrow the scope to specific paradigms. \cite{surveyN5} is the first comprehensive survey dedicated to energy efficiency in FL. It reviews strategies such as client selection, communication-efficient aggregation, compression and adaptive scheduling, and discusses how FL can be integrated with enablers like blockchain and 6G edge networks. Importantly, it also identifies tradeoffs between energy efficiency, model accuracy, and convergence speed. However, the survey remains limited to the FL paradigm and does not generalize its findings to other 6G scenarios such as unmanned aerial vehicle (UAV) communication, Reconfigurable Intelligent Surface (RIS) or vehicular networking.

In a similar line, \cite{surveyN8} proposes a multilayer heterogeneous network architecture called MUSIC for energy-efficient distributed machine learning. It shows how ML tasks can be split across device, edge and core layers to reduce energy consumption and incorporates mechanisms such as Device-to-Device (D2D) collaboration and multiple access schemes. This survey provides concrete architectural insights but is focused on a single solution space, without extending the discussion to other AI paradigms or use cases.

Finally, \cite{surveyN6} presents a more model-centric view by categorizing AI methods for green communications as heuristics, classics in ML and deep learning. It discusses how the models can aid cellular communications, machine-type communications, and computation-based communications in the forthcoming 6G network. It points out the enabling technology like UAVs, RIS, SAGINs and energy harvesting, and open issues like the computational overhead of AI models along with the need for lightweight designs. Whereas it provides a systematic taxonomy of AI models towards EE, a unifying framework that can be used as a basis for comparing their adaptability towards varied 6G dynamics is missing. 

Collectively, these AI-for-EE surveys confirm that energy efficiency is a recognized and innovative line of work in 6G. They offer either high-level mappings of AI approaches to EE objectives or in-depth explorations of particular paradigms like FL or distributed ML. Nonetheless, they still remain disconnected: none of them organize such EE strategies across 6G use cases, and none methodologically study the way AI approaches accommodate the intrinsic dynamics of 6G. Filling this gap, our survey integrates heterogeneous EE strategies across a number of representative use cases and examines AI adaptability and EE tradeoffs on the fly in dynamic 6G settings.

\subsection{Domain-specific surveys (partial EE focus)}
A second group of surveys focus on the role of AI in energy optimization within particular verticals of the 6G ecosystem, such as IoT, industrial systems, vehicular networks, non-terrestrial networks or Wi-Fi. By focusing on a single domain, these works provide valuable depth and context-specific insights, but their findings often remain specified to that vertical, limiting broader generalization.

In the IoT scenario, \cite{survey6} surveys resource optimization approaches with a focus on computation offloading, energy harvesting and spectrum allocation. The survey underscores the potential for AI assistance in urgency prediction, network load modeling and adaptive service control in constrained IoT settings. This presents a practical outlook on edge intelligence for battery-powered devices, but the discussion is not structured around AI decision objectives and does not address the adaptation under non-stationary traffic or mobility. In addition, authors in \cite{survey2} survey resource management of LTE, 5G, and 6G comparing approaches such as network-aware dynamic allocation, predictive allocation according to traffic patterns and topology and energy-efficient allocation leveraging device location and mobility. The article encompasses a wide range of applications including Intelligent Transportation Systems (ITS), industrial internet of things (IIOT) and Mobile Crowdsensing (MCS) and discusses future research directions. However, AI is briefly noted in selected approaches and is not taken as the overarching unifying methodology. Expanding towards industrial settings, \cite{survey11} focuses on energy-efficient IIOT under the green 6G networks. It integrates enabling technologies like RIS, UAVs and D2Ds and stresses edge AI, federated learning and digital twins for sustainable factory settings. Nevertheless, the analysis is heavily architectural and fails to describe AI models as well as adaptability towards heterogeneous dynamics.

Switching to city-scale views, \cite{survey7} surveys the smart infrastructure scenarios such as smart grids, urban services, and Industry~5.0. It showcases decentralized, data-based control towards sustainability and resilience and sets AI as a facilitative enabler. However, the analysis is still vision-based without systematic discussion of specific AI models, quantification of EE gains or tradeoffs under time-varying demand. For connectivity in hard-to-reach regions, \cite{survey8} surveys non-terrestrial networks (satellites and UAVs) towards energy-efficient IoT. It is oriented towards hardware and protocol adaptations under harsh power constraints without any specific reference to AI. The survey does not evaluate whether learning-based control is still robust under the large delay and intermittent links characteristic of NTNs.

In latency-constrained automotive applications, \cite{survey9} considers approximate edge AI for self-driving cars, comparing compression methods, early-exit models and inference scheduling to balance accuracy and energy. It gives detailed technical discussion but is limited to vehicular environments. Expanding on this, \cite{surveyN7} directly characterizes the tradeoff between energy efficiency and quality of experience(QoE) in vehicular networks, following the development from the 4G through the 6G generations and pointing out industrial vehicular and wise transportation use cases. It sets the scene for AI as the way forward to trade off human-centric QoE and sustainability goals but is again limited to the vehicular vertical. 

Outside the domain of cellular, \cite{surveyG} compares the path towards AI/ML-native Wi-Fi with a focus on spectrum access and interference control in IEEE~802.11. Although it provides a clear domain-specific example, energy efficiency is not a keypoint and the extension towards broader 6G ecosystem is weak. 

In total, these domain-specific surveys clarify the ways AI can facilitate energy-aware operation in IoT, IIoT, vehicular, NTN and Wi-Fi contexts. They offer vertical depth and point out context-dependent challenges, yet they do not explore cross-use-case AI patterns, categorize methods by decision goals or systematically analyze adaptability under heterogeneous 6G dynamics. Our survey fills the gap by presenting a common cross-domain synthesis that translates AI methods into specific energy goals, assesses their adaptability with respect to the key dynamics of 6G and defines the associated tradeoffs.

\subsection{Vision and framework-oriented surveys}
Another strand of the literature does not attempt to provide exhaustive taxonomies of methods or general panoramas of 6G, instead it advances conceptual frameworks and visions where AI is positioned as a foundational element of future wireless systems. These works are characterized by the introduction of explicit models, paradigms or architectures that aim to guide research and development toward sustainable and intelligent 6G.

\cite{survey12} presents the Pervasive Multi-Level AI (PML-AI) framework as a combination of distributed agents, knowledge graphs, and digital twins under a layered 6G framework. It is due to support real-time and non-real-time decision making with light-weight AI models at the edge and digital twins as part of the pre-validation before performing actions. Three key challenges the survey assumes as needed for sustainable AI in 6G are energy cost, latency and controllability. It suggests energy-saving mechanisms such as pruning, quantization, and feature compression. While fundamentally conceptual, this work is not practical in the sense that it does not test in reality how these methods cope with the dynamic conditions of the 6G including variability in traffic, heterogeneity in services or user mobility.

A complementary vision is provided by \cite{survey13}, which advocates for Wireless Big AI Models (wBAIMs) as a new paradigm tailored to the needs of wireless communication. It describes a development pipeline based on pretraining, fine-tuning and distillation and highlights potential benefits such as hyper-reliable communication, integration of sensing and communication and intelligent orchestration of resources. The authors mentioned EE as a concern, particularly in relation to inference delay and computational complexity but the main contribution lies in presenting a paradigm shift toward large-scale wireless models rather than a detailed analysis of EE strategies.

From a sustainability perspective, \cite{surveyN1} reframes the shift from efficiency towards sustainability in 6G. It overviews the enabling mechanisms such as energy-aware design, green base stations, renewable integration and edge/cloud assistance and relates them to the overall Sustainable Development Goals via the suggested "quintuple helix" framework, which entails academia, industry, government, society and environment. This paper is unique in advancing the AI–EE discourse towards ecological and policy paradigms but it does not delve into the specifics of AI models or their adaptability to technical dynamics in 6G.

On the same note, \cite{surveyN3} attempts to bridge the transition from 5G to 6G with a particular focus on energy, IoT and ML. It surveys over 370 works, covering system designs, enabler technologies and autonomous networks and smart grids among others. Its added value is scope and audacity: the work demonstrates the convergence across energy considerations, IoT expansion and the abilities of ML in the determination of the 6G. Although it points out the challenges and possibilities, no systematic taxonomy across AI methods is given nor adaptability given towards dynamic conditions in the 6G.

Last but not least, \cite{surveyB} suggests a 6G architecture that is layered into a physical network layer, a slicing layer, a native intelligence layer, and a security foundation. It suggests the importance of AI in all the layers and points towards orchestration mechanisms towards networking, sensing and computing integration. The architectural approach is helpful in thinking about the 6G system as a system of systems but too general in handling AI without particular reference to energy-reducing mechanisms or adaptability. 

In general, these framework and vision-based works provide reference models and conceptual roadmaps that connect AI with the sustainable design of the 6G. They are innovative as they put forward paradigms and frameworks that can guide the future work. Although they are high-level, systematic studies on tangible AI models, on their energy-saving possibilities and on their feasibility towards the heterogeneous and dynamic conditions that are anticipated in the 6G are missing.

\subsection{General 6G surveys with broad scope}
A final set of surveys is focusing on presenting comprehensive overviews on 6G technologies, applications, and enabling approaches. Their goal is to map the wide landscape of future wireless systems while discussing of spectrum, architectures intelligence, and security. While these works are valuable as reference points and for contextualizing research directions, they generally treat energy efficiency only peripherally.

\cite{survey1} surveys the application of AI and ML in IoT towards a 6G vision indicating how various algorithms could enhance sensing, communication, and management operations. Even though the survey inspects IoT-AI synergies, this survey fails to frame energy efficiency as the key metric nor evaluate model adaptability in non-static environments. In the same manner, \cite{survey3} touches on the convergence between AI and the 6G referring towards the need for intelligence embedding towards self-autonomous as well as adaptive networks. Treatment is again high-level and conceptual without systematic considerations on EE goals.

Among the most ambitious perspectives, \cite{survey10} frames the 6G as the "Intelligent Network of Everything". It weaves wireless communication, AI and Internet of Everything into a unified narrative and addresses enabling technologies like RIS, THz, federated learning and semantic communication. Though energy and sustainability are mentioned among the performance targets, the paper fails to evaluate AI models towards EE or investigate their adaptability under few specific aspects. \cite{surveyA} equally sketches the roadmap by categorizing emerging use cases into distinct groups. The importance of AI in handling heterogeneity and large-scale data is stressed but energy concerns are touched upon briefly without technical elaboration on EE strategies.

Other works have more limited scopes. \cite{surveyD} is a collection of articles on communication system deploying machine learning with data-driven channel estimation and resource allocation solutions but without providing a unified taxonomy or 6G direction. \cite{surveyE} surveys ML use cases throughout the wireless stack and QoS and QoE optimization insights but lacks coverage of EE or adaptability under certain 6G-specific environments. Lastly, \cite{surveyF} is a survey on AI-blockchain integration towards wireless network trust and performance. While the perspective is original, no consideration is given to energy efficiency nor is the energy-constrained setting analyzed under AI models. 

In short, these wide-scope surveys are helpful panoramas of 6G and show the heterogeneous collection of enabling technologies that will be used in next-generation networks. Their strength lies in contextual coverage and horizon scanning but they remain general in orientation: they do not put AI-driven energy efficiency at the core nor give a framework that can be taken to analyze adaptability into the dynamic and heterogeneous conditions that will be found in 6G. Our survey goes beyond these works by putting AI-for-EE centerstage and by formally evaluating adaptability and tradeoffs through specific 6G use cases.

\begin{table}[htbp]
\centering
\caption{Classification of existing surveys according to their main scope}
\label{tab:survey-categories}
\begin{tabular}{p{4cm} p{4cm}}
\hline
\textbf{Category} & \textbf{Surveys} \\
\hline
Dedicated AI-for-EE surveys &
\cite{survey5}, \cite{surveyN5}, \cite{surveyN6}, \cite{surveyN8} \\

Vision and framework-oriented surveys &
\cite{survey12}, \cite{survey13}, \cite{surveyN1}, \cite{surveyN3}, \cite{surveyB} \\

Domain-specific surveys (partial EE focus) &
\cite{survey6}, \cite{survey2}, \cite{survey11}, \cite{survey7}, \cite{survey8}, \cite{survey9}, \cite{surveyN7}, \cite{surveyG} \\

General 6G surveys with broad scope &
\cite{survey1}, \cite{survey3}, \cite{survey10}, \cite{surveyA}, \cite{surveyD}, \cite{surveyE}, \cite{surveyF} \\
\hline
\end{tabular}
\end{table}

\subsection{Positioning of this survey and contributions}
Even though very recent literature has explored the convergence between AI and 6G, the majority of the existing surveys remain either too broad-based or too specialized to capture the specific problem of energy efficiency under dynamic conditions in the 6G scenario. Broad surveys like \cite{survey1}, \cite{survey2} and \cite{survey10} offer high-level overviews on the enabling technologies but fail to scrutinize the latter's role in terms of energy-aware decision handling under the hard real-time constraints. Taxonomy-based works like \cite{survey5} and \cite{survey11} categorize the AI techniques and the green enablers and refrain from making assessments on the adaptability of the models across mobility, traffic variability or service heterogeneity. Architecture-oriented visions like \cite{survey12}, \cite{survey13} and \cite{surveyB} offer some useful frameworks for the distributed intelligence and sustainable design of the 6G network but remain conceptual and fail to make specific analyses on the AI-driven EE strategies. Vertical-specific surveys \cite{survey6}, \cite{survey8} or \cite{survey9} examine IoT, non terrestrial networks or vehicular communications in detail but their insights remain limited to isolated verticals without synthesizing lessons across the broader 6G ecosystem. Even works on the AI-native wireless stack as well as the trust-based framework works do not map between the AI goals and the energy optimization goals considering the dynamic conditions in the 6G scenario.

To note such gaps, we have developed a comparative survey across all previous surveys listed in Table~\ref{tab:comparative-surveys}. It compares all surveys against eight criteria: 6G focus, EE focus, use-case orientation, classification of AI by functional goal, adaptability to 6G dynamics, explicit identification of dynamics, adaptability mapping and tradeoff analysis. This systematic comparison reveals that although previous surveys provide useful information, none do all eight in one go. Our survey is, however, the first one that incorporates all eight comprehensively.

The contributions of this survey can be summarized as follows:
\begin{itemize}
\item \textbf{use-cases oriented:} We recognize and align the topmost 6G scenarios in which EE is the paramount requirement such as RIS-assisted communication, UAV-based relaying, smart IIoT environments, V2X systems and adaptive resource management.

\item \textbf{AI-driven energy optimization strategies:} For each use case, we synthesize the state of the art of AI approaches addressing EE challenges covering reinforcement learning, federated learning, model compression, predictive orchestration and other interesting techniques.

\item \textbf{Adaptability analysis in 6G dynamics:} We introduce a unified framework of dynamic aspects and conduct a thorough evaluation of how well the reviewed AI solutions adapt to these evolving conditions. The assessment is carried out across three distinct timescales: real-time, proactive and static to capture all levels of adaptability and examine the corresponding resilience mechanisms.

\item \textbf{Visual synthesis across domains and visual comparison:} Structured tables and visual diagrams that align AI goals are built enabling agents and dynamic flexibility across all surveyed papers. They offer conceptual intuitiveness as well as practical design directions.

\item \textbf{Tradeoffs in energy efficiency analysis:} We recognize and interpret six key tradeoffs, including energy versus latency, fairness, coverage, computational complexity and accuracy. We demonstrate how AI methods can be customize to strike these conflicting objectives in dynamic 6G environments. \end{itemize}

In this fashion, our survey provides a comprehensive, technically-oriented and adaptability-conscious guide on AI-powered energy efficiency for 6G. It can be utilized as both: a consolidated research foundation and as a guide for future work on sustainable AI-native 6G systems.

\section{Energy efficiency and sustainability in 6G use cases: From standardized visions to EE-first lenses}
The discussion on the use cases of 6G is already in full swing inside standardization bodies and industrial consortia, and energy efficiency and sustainability always pop up as key themes. The ITU-R IMT-2030 framework~\cite{ituR_m2160} categorizes six families of usage scenarios: immersive communication, massive communication, low-latency hyper-reliable communication, ubiquitous connectivity, AI and communication and integrated sensing and communication. Differently from 5G, sustainability is treated by IT-U R as a cross-cutting design principle across all the scenarios, embedding energy efficiency as a key requirement and not as a sideline. In the same way, 3GPP has started Release-20 studies on the service requirements of 6G~\cite{3gpp_sa1_workshop2024} in which sustainability and energy efficiency show up as part of the Stage-1 service framework. Indeed, these initiatives confirm that energy sustainability will not only be limited to one vertical but also it has to be embedded into all service categories envisioned for 6G.

Industry and regional efforts support this view. The Next Generation Mobile Networks (NGMN) Alliance points environmental sustainability as a non-negotiable 6G design requirement~\cite{ngmn_2022} calling for AI-based mechanisms that decrease network power consumption as well as carbon footprint. The European Hexa-X and Hexa-X-II projects similarly put forward energy efficiency as the focal point outlining digital twins and edge/cloud orchestration as major enablers for end-to-end optimization~\cite{hexaX_d1.2,HexaXII-D2.6}. ETSI further pursues this goal through its ENI and NFV groups which specify energy-reducing use cases such as adaptive base station sleep and energy-aware orchestration as well as KPIs for the measurement of energy performance~\cite{etsi_eni001,etsi_nfv021}. Taken as a whole, these efforts all arrive at the same conclusion: energy efficiency and sustainability are not whiz-bang features but hard requirements that will define every facet of the 6G network.

\begin{figure}
   \centering
    \includegraphics[width=1\linewidth,height=0.9\textheight,keepaspectratio]{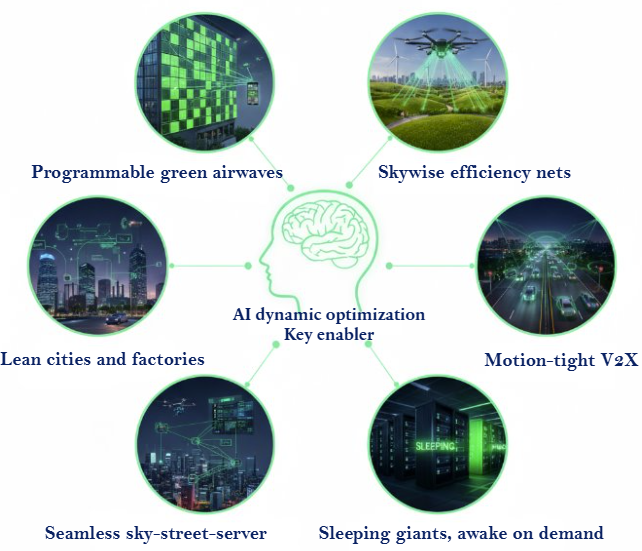}
    \caption{The 6G EE-first lenses}
    \label{fig:usecases}
\end{figure}

Based on these premises, our survey presents a new set of \textbf{energy-efficiency-first lenses} depicted in figure \ref{fig:usecases}. The use cases we propose reinterpret the standardized usage scenarios through the lens of sustainability. Whereas the service that 6G should provide is depicted by the prescriptions written by the ITU-R and the 3GPP, our approach revolves around the question of how such services can be provided in a sustainable way making AI the enabler of adaptive energy optimization. In order to make such perspective clear and organized, we restructure the standardized use cases into six EE-first use cases:
 
(1) \textit{Programmable green airwaves} (RIS-assisted communication),  
(2) \textit{Skywise efficiency nets} (UAV-based coverage),  
(3) \textit{Lean cities and factories} (IIoT and smart infrastructures),  
(4) \textit{Motion-tight V2X} (vehicular networks under high mobility),  
(5) \textit{Sleeping giants, awake on demand} (network-wide operations), and  
(6) \textit{Seamless sky–street–server} (cross-domain orchestration).  

The next subsections give a brief description on the purpose of each EE-first use case before detailing the technical design in later sections.
\FloatBarrier 

\subsubsection*{Programmable green airwaves (RIS-Assisted EE)}

The goal is to turn the wireless ambiance itself into a saving resource on the energy front. RIS enables the control of propagation without transmitting actively to decrease power at the base stations and end users. The challenge is to configure RIS in highly dynamic environments in a way that the spectral and energy efficiency is enhanced.

\subsubsection*{Skywise efficiency nets (UAV/Rotorcraft EE)}

UAVs are seen as agile relaying nodes and coverage points in 6G but propulsion energy leads their consumption profile. It is through this use case that the lifetime elongation of the UAV mission and the coverage reliability is enabled through AI-based trajectory planning, hover strategies and cooperative communication control decisions that aid general energy consumption reduction.

\subsubsection*{Smart factories and lean cities (IIoT and Smart Infrastructure EE}

IIOT and smart-city deployments will involve ultra-dense sensors, autonomous cars and smart infrastructures. The objective here is to achieve sustainable running through the use of AI in predictive device duty-cycling, energy-conscious task offloading as well as digital twin-based planning. The challenge is to decrease power usage at the expense of vital service reliability.

\subsubsection*{Motion-tight V2X (Vehicular EE under high mobility)} Vehicular networks are high-speed mobile, latency-constraint and safety-critical communication. The challenge in this application is maintaing reliable V2X communication under low delay requirements with optimal energy consumption spent on transmission, reception and cooperative awareness.

\subsubsection*{Sleeping giants, awake on demand (Network-Wide EE Operations)} Future 6G networks will be dependent on ultra-dense base station and access point deployments. In order to prevent energy expensive costs, the use case attempts to reduce idle power consumption by smart activation and deactivation of network components. AI-based traffic predication and multi-agent control enables the infrastructure to go toward deep sleep when no longer needed without sacrificing service continuity.

\subsubsection*{Seamless sky–street–server (Cross-Domain Green Orchestration)} This use case envisions a comprehensive end-to-end energy sustainability across diverse domains. By coordinating communication, computing and sensing across terrestrial, airborne and non-terrestrial networks, AI can facilitate combined optimization of function location, workload allocation and routing in order to keep the total energy cost low. It should ensure global efficiency without compromising latency or coverage.

The following sections of the present survey detail these six use cases characteristic of EE, each of which synthesizes how AI methods are employed for energy optimization. For each use case, we evaluate the extent to which AI-based solutions can adapt to the inherent dynamics of 6G: mobility, channel variability, traffic volatility, topological changes, service heterogeneity, and resource constraints.

\section{Why is energy efficiency crucial for 6G network?}

As 6G networks evolve to support massive-scale connectivity, ultra-fast data rates and intelligent automation, energy efficiency emerges as a critical design requirement. Unlike previous generations, 6G is expected to integrate terrestrial, aerial and space-based infrastructures facilitating seamless global connectivity while ensuring minimal environmental impact. Achieving carbon neutrality and reducing power consumption across network components ranging from core infrastructure and edge computing to user devices and IoT sensors are key objectives in 6G's sustainability roadmap.

However, energy optimization in 6G is far more complex than in prior generations due to its dynamic and adaptive nature. Traditional energy saving mechanisms which rely on fixed power control strategies will no longer be sufficient. Instead, 6G networks must operate in highly variable environments, where real-time adjustments are required to balance performance, latency and energy efficiency. The shift to AI-driven self-optimizing networks is therefore essential for maintaining sustainability while delivering next-generation services.

The next section examines the key dynamic factors that influence energy consumption in 6G networks, highlighting the challenges that require intelligent adaptive energy management strategies.

\section{Dynamic aspects of 6G networks affecting energy efficiency}

\begin{figure}
   \centering
    \includegraphics[width=1.1\linewidth,height=0.33\textheight,keepaspectratio]{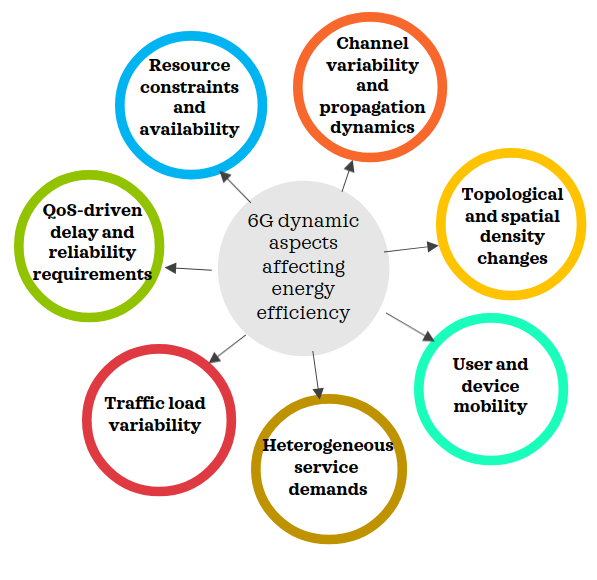}
    \caption{The inherent 6G dynamic aspects affecting EE}
    \label{fig:6g_dynamics}
\end{figure}

Energy consumption activity of the 6G networks is controlled by a group of dynamic factors depicted in figure \ref{fig:6g_dynamics} that arise from their unwonted heterogeneity, scale and flexibility. Such dynamics are significantly more volatile compared to the past wireless generations given the integration of aerial platforms, satellite relays, reconfigurable surfaces, mobile edge computing and ultra-dense IoT ecosystems. Consequently, predefined or static energy control policies are no longer viable. Instead, the intelligent systems should be capable of adapting smoothly under fast-changing environmental and service level conditions.

\paragraph{User and device mobility.}
In 6G settings with UAVs, autonomous vehicles and mobile IoT devices, dynamic positioning creates regular topology changes. Energy efficiency depends on the ability of the system to adjust trajectories, handovers and scheduling to maintain link quality while minimizing movement and signaling overhead.

\paragraph{Channel variability and propagation dynamics.}

High-frequency bands used in 6G (e.g., THz, mmWave) suffer from rapid fading, path blockage and directional instability. Adaptive energy control requires learning-based beamforming, RIS tuning or modulation decisions that can respond to unpredictable signal conditions.

\paragraph{QoS-driven delay and reliability requirements.}
Energy-saving strategies such as deep sleep modes can conflict with URLLC constraints. AI agents must continuously balance delay tolerance with energy cost often through shaping of dynamic reward or adjusting predictive policy.

\paragraph{Traffic load variability.}

Sudden spikes in traffic load affect buffer occupancy, transmission scheduling and mode switching. Intelligent systems must predict traffic behavior to enable preemptive sleep transitions or bandwidth reassignment without incurring delay or power penalties.

\paragraph{Resource constraints and availability.} 6G networks must operate under strict limitations in energy supply, computational capacity and spectrum allocation especially for battery-powered devices. These constraints require adaptive power control, lightweight AI inference and context-aware offloading strategies that minimize energy waste while preserving task utility.

\paragraph{Topological and spatial density changes.} In contrast with fixed deployments, 6G incorporates time-varying topologies via mobile UAVs, shifting RIS and dynamic clustering. They dynamically recalculate the communication graph as well as energy paths imposing AI models that can re-learn the optimal configuration with low retraining overhead. 

\paragraph{Heterogeneous service demands.} 6G must concurrently support eMBB, mMTC and URLLC. These services impose conflicting constraints on latency, reliability and energy use. AI-driven solutions must learn how to dynamically prioritize energy resources across services while minimizing QoS degradation. 

These representative dynamics show why energy optimization in 6G should be considered as a learning-based and context-aware process. Table~\ref{tab:dynamics} summarizes this collection of dynamic factors that impact energy efficiency along with definitions, examples in the real world and the AI adaptation challenges they impose.

\FloatBarrier 

\begin{table*}[t]
\centering
\caption{Key 6G dynamic aspects affecting energy efficiency and AI adaptability}
\rowcolors{1}{green!12!white}{green!2!white}
\arrayrulecolor{black}        
\setlength{\arrayrulewidth}{1pt} 
\renewcommand{\arraystretch}{1.25}
\small 
\begin{tabular}{|p{2.5cm}|p{4.5cm}|p{3.5cm}|p{4.5cm}|}
\hline
\textbf{Dynamic Factor} & \textbf{Definition} & \textbf{Example} & \textbf{AI Adaptation Challenge} \\
\hline
\textbf{User and device mobility }& Continuous movement of UAVs, vehicles or users affecting topology and link quality & UAV trajectory updates in aerial coverage scenarios & Real-time trajectory or handover optimization without full retraining. \\
\hline
\textbf{Channel variability and propagation dynamics} & Fluctuations in signal quality due to fading, blockage and directional instability & LoS blockage in mmWave UAV-to-ground links & Adaptive beamforming, RIS tuning and power modulation under fading. \\
\hline
\textbf{Traffic load variability} & Temporal variation in data traffic demand across users and services & Burst traffic from event-triggered sensor clusters & Predictive load adaptation and sleep/wake mode scheduling. \\
\hline
\textbf{Resource~constraints and availability} & Limitations in energy, bandwidth, or computation on nodes or edge devices & Battery-powered IoT sensor nodes with bursty load & Energy-aware task offloading and AI model quantization under constraints. \\
\hline
\textbf{QoS-driven delay and reliability requirements} & Latency, reliability, and throughput demands set by the service-level agreement (SLA) & Delay budget for URLLC and real-time AR/VR links & Balancing energy-saving actions with hard QoS targets via reward shaping. \\
\hline
\textbf{Topological and spatial density changes} & Reconfigurable network due to mobile UAVs, RIS or clustering dynamics & UAV relocation or RIS repositioning in urban scenarios & Continuous environment sensing and policy updates for energy optimization. \\
\hline
\textbf{Heterogeneous Demands} & Coexistence of URLLC, mMTC, eMBB with diverse requirements & Simultaneous sensor reporting and HD streaming & Learning to balance energy usage across conflicting service profiles. \\
\hline
\end{tabular}
\label{tab:dynamics}
\end{table*}

\section{AI-driven solutions for energy efficiency in 6G networks}

Based on the identified dynamic aspects from the previous section, this part examines how artificial intelligence has been utilized to maximize the energy efficiency of 6G networks. Closed-loop adaptability is offered by AI-based solutions so that decisions are made in real time on multiple factors such as resource allocation, beamforming, mode selection and offloading that are directly sensitive to the various dynamics. In what follows, we survey the representative categories of AI-based energy-efficient solutions showing the associated technical mechanisms as well as the way the methods are adaptable to the fundamental 6G dynamics discussed above.

\subsection{AI for RIS-assisted communication}

RIS are programmable metasurfaces composed of many sub-wavelength “meta-atoms” that impose controllable phase/amplitude shifts on electromagnetic waves in a way that it can shape the radio channel itself with very low power overhead \cite{RIS00,RIS01}. Compared to large-scale MIMO beamforming that requires many RF chains, RIS offers a lightweight way to enhance coverage and link budgets especially at mmWave/THz by steering or focusing reflections with near-passive hardware.

RIS thus emerges as a promising enabler of energy-efficient 6G communication. By dynamically controlling how electromagnetic waves reflect, RIS allows the wireless environment itself to become programmable, creating energy-efficient propagation paths, reducing transmit power and mitigating blockage without deploying additional active radio chains. Yet, the effectiveness of RIS-assisted communication depends on the real-time optimization of reflection coefficients under high-dimensional, dynamic and partially observed conditions making it a natural candidate for AI-based control.

The integration of AI in RIS-assisted 6G systems enables the RIS controller to learn adaptive configurations of reflection phase shifts and gains in response to user location, channel state, interference and service demand. When RISs are mobile such as those mounted on UAVs or operate in multi-hop relays, the complexity increases with UAV trajectory, user association, beamforming and RIS configuration, quickly making traditional optimization infeasible. In contrast, reinforcement learning and deep learning approaches provide scalable and model-free alternatives capable of online learning.

Several works illustrate this direction. For instance, \cite{RIS1} addresses the energy efficiency optimization problem in multi-access edge computing (MEC) IoT networks by introducing an active flying RIS (FRIS) mounted on a UAV. Unlike conventional passive RIS, the active RIS can both reflect and amplify incident signals thereby alleviating the severe "double path-loss" effect and extending coverage in obstructed urban areas. The system jointly considers UAV trajectory planning, FRIS reflection optimization, IoT device offloading and MEC resource allocation, and employs a PPO-based DRL algorithm PPO-RATDFRO: (proximal policy optimization with resource allocation, trajectory design, and FRIS reflection optimization) to optimize these parameters in real time.

Similarly, \cite{RIS2} addresses the challenge of energy efficiency in RIS-assisted D2D communications within 6G smart city environments. D2D offers low-latency local connectivity but suffers from severe co-channel and cross-channel interference, especially under ultra-massive connectivity. To mitigate this, the authors integrate UAVs as aerial relays, Non-Orthogonal Multiple Access (NOMA) for spectrum efficiency and RIS to enhance propagation without excessive power use. The joint optimization of subchannel assignment, power allocation, RIS phase shifts and user association is formulated to maximize overall energy efficiency under QoS constraints for both cellular users and D2D pairs. To fulfill this goal, the authors design a multi-agent deep reinforcement learning (DRL) framework. Specifically, they propose the PS-DC-DDQN algorithm (Priority Sampling–Decentralized and Coordinated–Dueling Deep Q-Network). In this design, each UAV and D2D transmitter acts as an agent that learns local policies while exchanging minimal information with neighbors, thereby reducing communication overhead. Resource and power allocation are handled in a distributed yet coordinated manner while RIS phase optimization is addressed centrally through a C-DDQN scheme. The decentralized structure ensures scalability in dense networks while the RIS optimization reduces UAV power consumption by establishing virtual line-of-sight links. The study also demonstrates that the framework meets user QoS requirements more reliably than competing methods under high mobility and interference conditions.

\cite{RIS4} considers integrated satellite–aerial–terrestrial relay networks (ISATRNs) as a candidate architecture for next-generation massive IoT access where a high-altitude platform (HAP) acts as a central relay and UAVs equipped with RIS enhance ground coverage. To improve spectral/energy efficiency, the system adopts a mixed FSO/RF transmission mode: FSO for satellite–HAP downlink and RF for HAP–IoT ground links augmented by NOMA for spectrum efficiency. The motivation is to leverage RIS-equipped UAVs for energy-efficient coverage extension in dense IoT deployments under stringent UAV energy constraints.
The authors formulate a joint optimization problem involving UAV 3D trajectory, HAP active beamforming and RIS passive reflection matrices in order to maximize the system ergodic sum rate under UAV power consumption limits. To address this, they design an long short-term memory~(LSTM)-double deep Q-network~(double-DQN) algorithm. The LSTM module captures long-term temporal dependencies in UAV trajectory and channel evolution while double-DQN mitigates Q-value overestimation and enhances stability. A prioritized experience replay strategy is incorporated to accelerate convergence by focusing on more informative exploration samples.
LSTM-double-DQN with NOMA achieves the highest ergodic rate and energy efficiency of the system and outperforms classical double-DQN by learning more efficient UAV trajectories with fewer redundant paths and higher quality of service for distant users. The UAV has been shown to save energy by avoiding unnecessary detours and the RIS helps reduce the transmit power of the HAP while ensuring the QoS of the user.

Beyond coverage and connectivity, RIS has also been applied to enhance secure and sustainable communications. \cite{RIS5} addresses the dual challenge of safeguarding sensitive transmissions and minimizing power consumption in RIS-assisted multi-user wireless networks with multiple eavesdroppers. The system jointly optimizes transmitter beamforming, artificial noise vectors and RIS phase shifts with the aim of maximizing secrecy energy efficiency (SEE). This latter is defined as the ratio of secrecy sum rate to total power consumption. A deep deterministic policy gradient (DDPG)-based framework is proposed that uses actor-critic networks, replay buffers and target networks to manage continuous action spaces. The design explicitly integrates penalties for violating user rate requirements or exceeding transmit power budgets ensuring that the learned policies balance secrecy and energy efficiency under dynamic network conditions. By jointly encoding secrecy as well as energy requirements, this work departs from the traditional secrecy rate maximization approach highlighting the potential of DRL achieving secure as well as energy-conscientious connectivity within the context of RIS-assisted 6G.

Ahmad et al. in \cite{RIS6} address electromagnetic interference (EMI) from Gallium nitride (GaN) power amplifiers, often overlooked in UAV–RIS systems. They propose a DRL-driven framework integrating UAV trajectory control and RIS phase optimization with QPSK modulation to mitigate EMI while enhancing energy efficiency. Evaluated under urban, suburban and rural Rician fading, the method achieves an improvement of up to 6.5 dB SINR, a gain of 38\% EE, and a coverage extension of 35\% compared to the baseline UAV-RIS schemes. The approach demonstrates strong resilience to EMI but raises scalability challenges beyond 256 RIS elements and robustness concerns at high transmission power levels.

Finally, \cite{RIS7} investigates the integration of RIS with NOMA to enhance spectral and energy efficiency in dense IoT networks. The authors first introduce a user clustering strategy that groups devices with strong and weak channel disparities to mitigate interference and improve fairness. The joint optimization of RIS phase shifts and base station power allocation is then formulated to maximize both sum rate and energy efficiency, initially addressed with fractional programming and Karush–Kuhn–Tucker (KKT) conditions. To address the scalability limits of iterative optimization, two AI-based solutions are proposed: an algorithm Environment-Driven Deep Learning (EDDL) which learns phase power policies online without requiring large training data sets and an algorithm Exploration-Attenuated Deep Deterministic Policy Gradient (EA-DDPG) which handles continuous state/action spaces for long-term reward optimization. Simulation results confirm that both AI-based approaches significantly outperform baselines with EDDL offering faster convergence and EA-DDPG achieving higher steady-state energy efficiency.

Overall, these studies highlight the diverse roles of AI in unlocking the potential of RIS for energy efficiency in 6G. Across UAV-assisted, D2D, satellite–aerial–terrestrial, IoT and secure communication scenarios, DRL consistently emerges as a key enabler to solve non-convex optimization problems demonstrating superior adaptability compared with iterative methods. 

Across these works, AI contributes to RIS adaptability with respect to several core 6G dynamics:

\begin{itemize}
  \item \textbf{Channel variability and propagation dynamics} especially in mmWave and THz environments prone to blockage;
  \item \textbf{Topological and spatial density changes} such as RIS-equipped UAVs or mobile surfaces;
  \item \textbf{User and device mobility} which continuously changes optimal reflection angles and coverage areas;
\end{itemize}
In summary, AI techniques enable RIS-assisted communication systems to autonomously adapt to highly dynamic wireless environments and to unlock their full energy-saving potential within future 6G architectures.

\subsection{AI for energy-efficient industrial IoT and smart city systems}

IIOT and smart city networks are among the most demanding 6G use cases in terms of energy efficiency. These environments characterize ultra-dense device deployments, dynamic traffic profiles and mixed-criticality services ranging from delay-tolerant environmental monitoring to ultra-reliable control of robots, vehicles and infrastructure. Ensuring energy sustainability across such diverse deployments requires AI-driven techniques that span from low-level device control to high-level orchestration and deployment planning:

\subsubsection*{A) AI for energy-efficient device-level sensing and communication}
At the device level, AI is used to optimize energy usage through predictive scheduling, mode switching and transmission suppression. Instead of transmitting all sensed data blindly, sensors can use lightweight ML models to determine when communication is necessary and which mode (active, sleep, low-power) to operate in. For example, in \cite{paper87}, a hierarchical AI architecture is proposed in which clustered IoT groups are assigned using either convolutional neural network~(CNN) or backpropagation neural network (BPNN) models to predict latency, energy and performance. These predictions are used to selectively schedule transmissions, preventing unnecessary energy consumption in low-utility periods. In \cite{paper92}, authors train Random Forest and Decision Tree classifiers to predict energy consumption patterns of individual sensors and classify their operational status. This enables dynamic energy-aware communication across a wide sensor grid.

\subsubsection*{B) AI-enabled orchestration, mobility, and QoE-aware service management}

At the service level and infrastructure level, energy efficiency in smart city networks and IIoT is contingent upon the level of resources orchestrated between edge and cloud servers and cloud as well as the scaling and tailoring of such services to changing demand. By intelligently determining what servers, functions or communication paths are active each time, much larger energy savings can be achieved beyond the level of device-level savings. By itself, AI brings the predictive and adaptive intelligence necessary to make such orchestration possible under heterogeneous loads, strict Quality of Service requirements and mobile cases.

In \cite{paper44}, a multilayer edge–cloud architecture is conceived where AI modules predict service demand and traffic intensity to determine when and where to activate or deactivate edge servers. Clustering algorithms and supervised prediction models detect high-demand zones and anticipate load variations enabling dynamic allocation of edge resources. The system reduces the infrastructure's power footprint by aggressively putting inactive servers into low-power modes while scaling active servers only where necessary, without sacrificing responsiveness.
In addition to forecasting approaches, Wang et al. \cite{cc1} address the orchestration of service function chains (SFCs). They formulate orchestration as a joint energy–carbon optimization problem and propose a hierarchical deep Q-network with CNNs to manage SFC placement across large-scale IoT substrates. The hierarchical structure decouples global goal selection from detailed placement actions while CNNs extract features from network topologies. The approach reduces the number of active nodes as well as forwarding energy, thereby improving energy efficiency at scale.

Chen et al. \cite{cc2} tackle the same challenge from a distributed perspective focusing on parallel orchestration of VNFs and SFCs for edge intelligence in IIoT. They develop an Averaged Multistep Double-DQN (AMD2) algorithm that stabilizes and accelerates training while embedding VNFs as resource blocks across distributed servers. Minimizing the number of deployed VNFs and reducing transmission overhead leads the framework to decrease infrastructure energy consumption while sustaining throughput and service quality. Along with \cite{cc1}, this work is an illustration on how reinforcement learning is utilized to perform an energy-conscious service chain's orchestration.

Lastly, in highly mobile industrial settings, the orchestration must also account for mobility as well as QoE requirements. In \cite{paper17}, an architecture for 6G-enabled network-in-box~(NIB) is conceived where a mobility ML based management method (MMMM) is utilized for forecasting handover requirements, managing duty cycles and dynamically adjusting transmit power. This ensures energy-efficient connectivity for mobile devices such as robots or wearable IoT while meeting the strict QoS requirements of delay-sensitive services.

Combined, the works together point out the key role of AI in transforming orchestration from fixed provisioning to dynamic, energy-conscious management. Through the combination of forecasting, hierarchical and distributed reinforcement learning and mobility-conscious QoE optimization, the orchestration frameworks are able to minimize unnecessary activations, workload consolidation, and maintain service quality amidst spatiotemporal demand variability.

\subsubsection*{C) AI and digital twins for deployment planning in industrial environments}
Digital Twin (DT) technology \cite{dtsurvey,dtsurvey1,dtsurvey2} is emerging as an important enabler for sustainable 6G networks with virtual replicas of physical systems that allow proactive prediction, simulation, and optimization of network performance. By integrating DTs with AI methods, service providers are able to predict traffic changes, simulate work planning strategies and dynamically re-arrange resources, thus minimizing excessive energy consumption over industrial as well as smart city settings.

The work in \cite{paper45} presents a three-layer DT–AI architecture to support intelligent planning and sustainable operation of dense 6G industrial networks. The Physical Network Layer (PNL) collects heterogeneous factory data, the Twin Network Layer (TNL) performs situational awareness and decision making and the Network Application Layer (NAL) provides planning and optimization services throughout the network lifecycle. In fact, AI techniques such as clustering, supervised learning and DNNs are used for demand analysis and coverage optimization while reinforcement learning enables base station energy saving strategies and SDN-driven DRL supports load balancing. By combining simulation-driven planning with adaptive control, the framework minimizes trial-and-error costs and allows energy-aware deployment in smart factory environments.

Elmazi et al. \cite{dt1} extend the role of DTs to distributed intelligence in IoT sensor networks by combining DTs with FL. The DT, in this case, forecasts device-level variables like CPU usage, battery consumption and latency to develop an offloading decisions proactive layer. Based on this, the dueling double DQN agent makes decisions on the execution of tasks locally versus offloading them to the edge whereas FL takes care of mutual model training without the sharing of raw data. This joint DT–FL–RL framework allows devices to balance computation and communication in real time, thereby prolonging battery life and sustaining collective intelligence. The key contribution is the positioning of DTs as predictive enablers for energy-aware offloading and resource coordination in constrained IoT environments.

At the system level, Xu et al. \cite{dt3} put forward a holistic digital twin (DT) framework where AI is embedded throughout the lifecycle of modeling, simulation and optimization with energy efficiency as their clear design objective. Generative AI techniques are utilized to construct realistic virtual environments while DRL agents dynamically optimize base station and edge server. They switch policies to reduce overall energy consumption. This approach positions DTs as active, AI-driven platforms capable of anticipating network dynamics and executing proactive reconfigurations, moving beyond passive mirroring to system-wide orchestration for sustainable 6G.

Last but not least, Duran et al. \cite{dt4} focus on urban scale orchestration in 6G smart city IoT by introducing a DT-guided energy management framework. The architecture is organized into four layers data, twin, service and management. They are interconnected through the Real-Time Publish Subscribe (RTPS) protocol for low-latency updates. At the management layer, a DDPG agent determines optimal update schedules for IoT devices, selectively activating or deactivating nodes to preserve battery energy while sustaining service timeliness. By linking DTs to reinforcement learning engines, the framework transforms DTs into proactive controllers that reduce redundant device activity, enhance scalability as well as the energy footprint of large-scale smart city deployments.

Together, these AI-enabled approaches address a broad spectrum of 6G dynamics, including:
\begin{itemize}
  \item \textbf{Traffic load variability}, from predictive scheduling at the device level \cite{paper87,paper92} to edge–cloud orchestration \cite{paper44} and DT-based forecasting of demand \cite{paper45,dt3,dt4};
  \item \textbf{Resource constraints}, especially in energy-limited sensors and IoT devices \cite{paper87,paper92,dt1}, as well as in infrastructure nodes where RL-driven orchestration minimizes active servers and VNFs \cite{cc1,cc2};
  \item \textbf{Topological and and spatial density changes}, captured through mobility-aware orchestration in industrial NIB frameworks \cite{paper17} and DT-based simulation of factory or smart city layouts \cite{paper45,dt3,dt4};
\item \textbf{Heterogeneous service demands}, encompassing both delay-tolerant sensing and mission-critical control where user-perceived experience (QoE) is implicitly embedded in the diversity of service classes \cite{paper17,paper45,dt3};
  \item \textbf{QoS constraints}, which require energy-efficient orchestration under strict reliability, latency and continuity requirements that ensures sustainability without compromising service-level guarantees \cite{paper17,paper44,dt3,dt4}.
\end{itemize}

\subsection{AI for energy-efficient UAV communication}
UAVs are emerging as flexible communication platforms in 6G networks, providing on-demand coverage, relay assistance and sensing in a wide range of scenarios. Their role ranges from emergency response to industrial automation and rural access. However, UAVs are inherently energy-constrained not only in terms of communication power, but more critically in propulsion which dominates total energy consumption \cite{uav2,uav5}. To ensure sustainable operation of UAVs while providing reliable connectivity, AI techniques are increasingly used to optimize UAV trajectories, transmission strategies and coordination with ground or aerial infrastructure \cite{uav1,uav4}.

The general objective of AI in this domain is to guide the UAV’s movement and communication behavior in a way that maximizes coverage and service quality while minimizing propulsion energy and transmission overhead. Unlike traditional path planning, AI models can learn to balance multiple objectives in dynamic environments adapting to user mobility, wireless channel conditions, rate constraints or the presence of obstacles like buildings. Reinforcement learning, in particular, allows UAVs to explore the environment and learn policies that achieve energy efficiency without requiring explicit mobility or channel models.

Several works illustrate this direction. In \cite{paper36}, Q-learning is used to jointly optimize the altitude and transmission power of aerial base stations (ABSs) allowing UAVs to adjust power levels based on discrete height levels in clustered user regions. In \cite{paper100}, a Deep Q-Network is employed to determine UAV trajectories that minimize propulsion energy while satisfying rate outage constraints. The DQN learns optimal flying directions based on state observations and uses a dueling architecture to stabilize the value function. In \cite{RIS1}, proximal policy optimization (PPO) is applied in a RIS-assisted UAV-MEC scenario, where the agent learns to jointly control UAV movement and RIS configuration to enhance coverage and reduce energy cost under urban blockage conditions. Similarly, \cite{RIS4} extends this idea in a more complex setting involving satellite–HAP–UAV communication. An LSTM-enhanced double-DQN agent is used to adapt UAV trajectory in coordination with RIS phase tuning and HAP beamforming, enabling energy-efficient multi-hop relaying.

These UAV-based strategies are particularly responsive to several dynamic aspects of 6G:
\begin{itemize}
  \item \textbf{User and device mobility}, which alters coverage targets and optimal flight paths;
  \item \textbf{Channel variability and propagation dynamics}, such as NLoS conditions in urban canyons;
  \item \textbf{Topological changes}, including dynamic user clustering and repositioning of aerial assets;
  \item \textbf{QoS-driven constraints}, where delay, throughput, or outage duration requirements limit how aggressively energy can be saved.
\end{itemize}
AI techniques provide UAVs with a mechanism to continuously learn and adapt in these environments, achieving significant energy gains by jointly considering physical mobility and communication-layer dynamics.

\subsection{AI for energy-aware network operation and resource allocation}
One key strategy toward achieving the energy efficiency goal for 6G networks is intelligent management of network-level functions and resource allocation. This encompasses regulating base station (BS) activity, user association, spectrum reuse, load balance and spatial frequency planning so as to limit energy consumption while not compromising performance. AI techniques offer a scalable alternative by enabling context-aware, learning-based decisions that adapt in real time to environmental, traffic, and service-level changes.

The overall objective of AI within this application is to enable energy-conscious connectivity of the core and access network layers by smart management of the transmission power, BS on/off switching, user scheduling, as well as reusing resources. Instead of depending on deterministic models, the AI agents are trained from the feedback from the system so that they are capable of managing uncertainty, non-linearity, as well as dynamic QoS restrictions. These are deployed at the macro level most times, the idea being to conserve energy consumption throughout the overall network infrastructure, rather than of particular devices.

Illustrative works include \cite{paper98} where a State-Action Reward State-Action (SARSA)-based reinforcement learning agent is trained to manage the sleep states of the base station by observing the buffer load and traffic dynamics. The agent learns when to switch between active and sleep modes to balance energy consumption with QoS requirements such as queue stability and transmission delay. In \cite{paper37}, a multi-agent actor–critic framework is used to jointly optimize sub-band allocation, transmit power and power splitting ratios in a SWIPT (Simultaneous Wireless Information and Power Transfer) network. Each agent adapts its communication policy in a decentralized manner to ensure fairness and energy sustainability while considering spectrum reuse and interference. The work in \cite{RIS2} introduces a hybrid multi-agent DRL system where D2D transmitters optimize subcarrier and power allocation using local observations while a centralized DDQN agent configures the RIS to improve global energy efficiency across multiple users.

In addition, \cite{paper90a} proposes a hybrid quantum deep learning~(HQDL) framework that jointly optimizes network slicing, load balancing and base station energy states. A Base Station Optimizer Net integrates CNN-based traffic recognition and recurrent neural network (RNN)-based fault prediction to deactivate unnecessary base stations and efficiently reallocate slices. This multistage model reduces energy consumption while maintaining high accuracy in service continuity and congestion control. Similarly, \cite{paper47} presents a native AI simulation platform to evaluate MARL-based energy saving strategies in a two-tier 6G network. Using algorithms such as DDPG, COMA, and MAPPO, multiple agents collaboratively manage BS activity, communication resources, and task offloading decisions. Their coordination enables fine-grained power control and dynamic sleep mode activation based on real-time user demand and link quality.

The following 6G dynamics are particularly addressed in this category:
\begin{itemize}
  \item \textbf{Traffic load variability}, which affects buffer status, congestion levels and BS scheduling;
  \item \textbf{Resource constraints}, especially under strict energy budgets or dense access point deployments;
  \item \textbf{Topological and service heterogeneity}, including multi-slice architectures, overlapping coverage and adaptive spectrum sharing;
  \item \textbf{QoS-driven reliability and delay constraints}, which must be respected even as operational power is reduced.
\end{itemize}
By integrating AI agents across network control planes, these solutions transform traditional infrastructure into an adaptive, energy-aware platform that meets the ambitious scale of the 6G networking and sustainability goals.

\subsection{AI for energy-efficient V2X and high mobility scenarios}

Vehicular-to-Everything (V2X) communications represent one of the most energy-critical 6G use cases \cite{v1} with high requirements on latency, ultra-reliability and extreme mobility. Vehicles, roadside units (RSUs), and occasionally UAV relays are required to maintain stable links in extremely changing topologies under the constraint of limited energy supplies. Conventional optimization approaches for mode selection, resource allocation and power control struggle with the non-stationarity and large action spaces of such networks. To overcome these challenges, recent works leverage AI techniques primarily reinforcement learning (RL), multi agent reinforcement learning (MARL) and meta-learning to enable adaptive and real-time strategies for energy-efficient communication in high-mobility V2X environments.

\cite{paper16} suggests a mode selection scheme based on Q-leaning for underlay vehicular networks. Direct V2V mode versus cellular mode is dynamically selected by the vehicles to achieve maximum energy efficiency. The RL agent learns mode policies by maximizing vehicular downlink EE by up to 30\% over full D2D while ensuring fairness enhancement for cell-edge users. However, uplink EE for cellular users declines due to spectrum reuse revealing a trade-off that highlights the need for cross-layer interference coordination.

Building on this, \cite{v5} study V2X networks with RF-based energy harvesting, aiming to joint-channel selection, power allocation over the transmitter and harvested energy splitting so as to optimize EE under latency and SINR limitations. They design a decentralized multi-agent DRL framework based on PPO where each V2V transmitter simultaneously learns spectrum reuse, power allocation and energy division policies. Simulations show that the approach improves energy efficiency by about 30\% compared to MADQN and by over 150\% against fractional programming demonstrating robustness under mobility and channel variability.

Beyond EH-driven adaptation, authors in \cite{v6} address EE optimization in V2V networks with heterogeneous QoS requirements and partial CSI. They propose a distributed multi-agent parameterized DQN (DMA-PDQN) that jointly optimizes channel selection, transmission power and rate. To stabilize learning across agents, federated meta-learning is integrated enabling clustered agents to share knowledge while adapting locally. Results show that DMA-PDQN significantly outperforms DQN and PDQN discretization, improving EE and QoS satisfaction close to schemes with perfect CSI.

Scalability under dense topologies is also addressed by \cite{v3}. This work considers UAV-assisted V2X where UAVs act as relays under strict energy supply. The optimization of sub-band allocation, transmit power and UAV deployment is tackled via a multi-type mean-field MARL framework which approximates agent interactions by type to ensure scalability. Simulations show that the proposed scheme enhances system-wide energy efficiency and reduces link failure probability compared to DQN and MADDPG baselines especially in dense vehicular scenarios. The scheme also brings out the promise of mean-field MARL for scalable EE optimization, though UAV propulsive energy consumption and the coordination overhead are open issues.

To address continuous action spaces and non-stationarity, \cite{v4} combine twin-delayed DDPG (TD3) with dynamic meta-transfer learning (DMTL). This configuration enables rapid adaptation to new channel conditions. It achieves higher energy efficiency and link satisfaction probabilities with fewer training samples illustrating the promise of meta-learning in accelerating convergence.

Finally, \cite{v2} explore duplex DRL for joint TDD/FDD orchestration that aims to reduce energy waste at the radio access level while sustaining high throughput and reliability. Although EE gains are implicit, the approach demonstrates that duplex DRL can significantly improve resource utilization efficiency in dense V2X deployments.

As we can see in these studies, the central design AI patterns that enable energy efficient highly dynamic V2X networks are as follows: Q-learning for lightweight mode selection, multi-agent DRL for distributed power and channel allocation and meta-learning to accelerate adaptation under non-stationary channels. UAV relays and energy harvesting further extend the scope of EE-aware V2X but introduce new trade-offs such as propulsion costs and coordination overhead. Typically, small concessions in throughput or latency enable substantial energy savings, while implicit EE gains are often coupled to broader metrics such as reliability or fairness. Future directions call for integrated frameworks that account for communication and propulsion energy jointly, reduce coordination costs at scale and explicitly quantify trade-offs across mobility, reliability and energy objectives.

These AI-powered approaches address several core dynamic aspects of V2X and high mobility scenarios:
\begin{itemize}
  \item \textbf{User and device mobility}, with rapidly shifting link quality and association options;
  \item \textbf{Channel variability and propagation dynamics}, including fading, interference and NLoS conditions;
  \item \textbf{QoS-driven constraints}, especially ultra-low latency and high reliability for safety-related messages;
  \item \textbf{Topological and service heterogeneity}, as seen in mixed-mode (V2V, V2I) and multi-agent vehicular networks;
  \item \textbf{Resource constraints}, particularly energy-limited clusters or harvest-and-transmit V2X nodes.
\end{itemize}

\subsection{The transversal role of edge AI in energy optimization}

While this survey categorizes AI-driven energy efficiency solutions under key 6G use cases, it is worth noting the transversal role of edge computing that makes possible energy-aware intelligence across multiple domains. Rather than being passive compute offloaders, edge platforms actively host predictive models, distributed controllers and digital twin simulations that guide real-time energy optimization strategies~\cite{edgesurvey1,edgesurvey2}. For instance, in smart city and industrial IoT scenarios, AI modules in the edge predict service demand and dynamically wake-up edge nodes based on traffic density and energy profiles~\cite{paper44,edgesurvey3}. In smart factory environments equipped with digital twin~\cite{paper45,dt1,dt3,dt4}, edge computing facilitates localized training and inference reducing the need for energy-intensive backhaul communication. Similarly, cooperative edge servers participate in federated or distributed AI model updates to support energy-efficient learning across constrained IoT sensors~\cite{paper49,fledge1,fledge2,fledge3,fledge4}. Instead of isolating edge optimization into a dedicated category, this survey showcases its embedded impact across various use cases emphasizing its architectural significance toward scalable and energy-sustainable 6G networks.

\begin{sidewaysfigure*}[p]
  \centering
\includegraphics[width=1\linewidth,height=0.9\textheight,keepaspectratio]{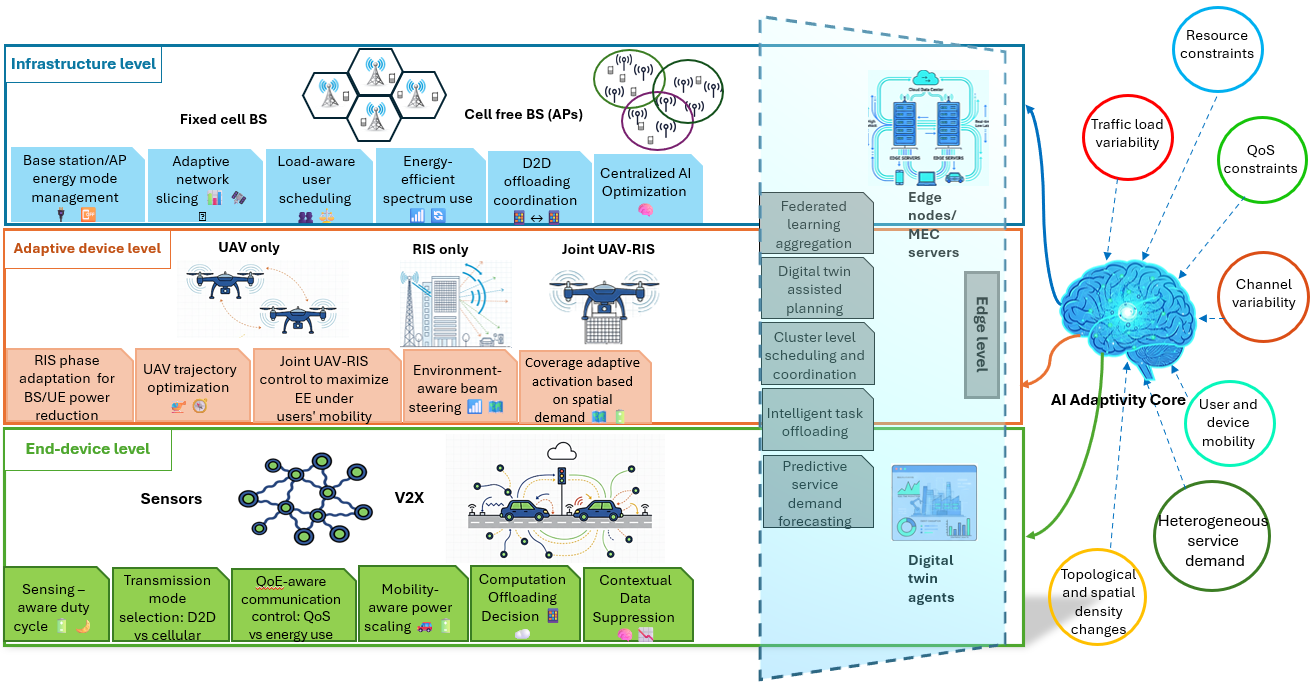}
    \caption{AI-driven energy efficiency objectives across functional levels in 6G networks. 
    Each horizontal colored box represents one main AI level: infrastructure, adaptive device and end-device. These three levels are crossed by one vertical edge box highlighting its crucial transversal role. Each box summarizes the main energy-saving functions and the associated physical or logical agents (e.g., fixed BSs, UAVs, MEC nodes). 
    On the right, the six core 6G dynamic aspects surround an AI adaptivity core (brain icon). 
    Dashed arrows indicate sensing/interpretation of dynamics; colored arrows denote dispatch of adaptive energy policies per layer.}
    \label{fig:ai_levels_dynamics}
\end{sidewaysfigure*}

To provide a unified perspective across the diverse AI-driven energy efficiency solutions we analyzed, we introduce a layered architectural perspective that organizes AI-driven energy efficiency strategies by their functional deployment within the 6G ecosystem. 
Figure~\ref{fig:ai_levels_dynamics} presents the three AI functional levels : infrastructure, adaptive device and end-device, each encapsulated in a colored module that summarizes its energy-related AI objectives and the physical or logical agents that implement them (e.g., fixed BSs, UAVs, MEC nodes). These modules are crossed by one vertical edge AI module to highlight its crucial transversal role. On the right, core dynamic aspects of 6G networks are visualized around an AI core illustrated in a brain icon with arrows representing how AI interprets these dynamics and dispatches adaptation policies to each level.

\FloatBarrier 

\section{Adaptability evaluation of AI-driven energy efficiency use cases to 6G network dynamics}

This section evaluates how well the AI-enabled EE strategies we've reviewed above respond to each of the core dynamic aspects. For every dynamic, we examine the AI techniques employed, their responsiveness across different use cases and the extent to which they address real-time variation and structural unpredictability.

\subsection{User and device mobility}

User and device mobility is one of the most defining challenges of 6G networks, particularly in use cases like V2X communication, UAV-assisted coverage, and mobile industrial IoT systems. High-speed movement, dynamic clustering and frequent topology changes require AI techniques that can make energy-efficient decisions under intermittent connectivity, frequent handovers,and dynamic resource reallocation. Adaptability here is not just about supporting mobile scenarios but about whether AI models can update policies quickly enough to preserve energy efficiency without degrading reliability.

Interestingly, RL methods have shown strong responsiveness to this dynamic since agents can continuously re-learn optimal actions from environment feedback rather than rely on static mobility models. In V2X systems, Q-learning and deep reinforcement learning approaches adapt mode selection, power control and spectrum reuse to rapidly changing vehicular topologies, although their performance remains sensitive to the quality of local state observations. More advanced methods such as mean-field MARL~\cite{v3} and meta-learning with transfer adaptation~\cite{v4} improve robustness by enabling scalable coordination across dense vehicle populations and rapid policy updates when mobility patterns change. Nevertheless, these methods incur training overheads and may lag in ultra-fast mobility contexts such as highway platoons.

In UAV-based networks, mobility is intrinsic to the platform. It makes the trajectory and the control of energy consumption highly dynamic optimization tasks. RL agents including Q-learning, PPO, and double-DQN~\cite{paper36,paper100,RIS1,RIS4} adjust UAV altitude, trajectory and transmit power in real time to maintain coverage while conserving propulsion energy. These solutions demonstrate genuine adaptability to mobility-driven topological shifts. However, they remain sensitive to partial or delayed channel state information limiting responsiveness under high-speed UAV movement or sudden blockage events.

In industrial and smart city contexts, AI solutions show more moderate responsiveness to mobility. Mobility-aware orchestration frameworks such as ML-based NIBs~\cite{paper17} predict handover events and adjust duty cycles, allowing energy-efficient communication for robots or wearable IoT. However, clustering-based prediction~\cite{paper87} or DT-based planning~\cite{paper45} tends to operate at slower timescales offering proactive but not real-time adaptation.

Overall, the reviewed AI techniques demonstrate strong adaptability to mobility when mobility-awareness is embedded into their learning loop. RL and MARL agents provide \textit{real-time} reactivity, meta-learning accelerates adaptation to unseen movement patterns with moderate overhead while DT-based frameworks and clustering-based prediction contribute at a \textit{proactive but slower} planning timescale. In contrast, static infrastructure-level methods only offer \textit{periodic} adjustments through retraining. This indicates that adaptability to mobility is strongest where online learning dominates, whereas proactive models complement by preparing long-term strategies but cannot handle abrupt changes.

\subsection{Channel variability and propagation dynamics}

6G environments will exhibit extreme channel variability due to diverse frequency bands, dense deployments and highly dynamic propagation conditions such as frequent transitions between Line-of-Sight (LoS) and Non-Line-of-Sight (NLoS), blockages and Doppler shifts. These fluctuations impact on signal quality, link reliability and the energy required to maintain continuous service particularly in UAV, V2X and RIS-assisted scenarios. 

RL methods, especially those that continue to learn through direct interaction with the environment rather than relying on pre-trained or model-based strategies, have shown strong potential for rapid adaptation when network conditions change. In UAV-based networks, DRL agents~\cite{paper100,RIS4} adjust flight paths, transmit power and beam directions to avoid obstructed or weak links maintaining energy efficiency under volatile propagation. Similarly, in RIS-assisted networks, DRL-based reflection optimization~\cite{RIS1,RIS2} adapts phase coefficients to channel observations reducing retransmission overhead and sustaining spectral efficiency. In V2X, meta-learning and mean-field MARL~\cite{v3,v4} further enhance responsiveness by allowing policies to generalize across fast-changing fading conditions and interference profiles. On the other hand, duplex DRL~\cite{v2} dynamically switches between TDD and FDD modes to handle uplink–downlink asymmetries caused by Doppler shifts.

At a slower but proactive timescale, digital twin frameworks~\cite{dt3,dt4} simulate channel propagation in advance. Thus, it enables base station switching and power scaling strategies in order to reduce energy waste under anticipated load or fading conditions. Although it is not suited to instantaneous channel changes, such predictive planning complements real-time RL by preparing longer-term reconfiguration strategies. In contrast, static optimization approaches and supervised classifiers that operate without channel-state feedback \cite{paper92,RIS1} can only adapt periodically or after retraining. As a result, they are generally poorly suited to environments where propagation conditions change rapidly.

In summary, the ability of AI to adapt to channel variability largely hinges on how well it incorporates channel-state information. RL and MARL techniques excel at reacting to instantaneous fading and blockage but remain limited by CSI acquisition overhead and scalability in dense deployments. Meta-learning improves responsiveness to unseen Doppler or fading conditions whereas DT-based approaches provide proactive foresight at slower timescales but cannot react in real time. RIS optimization currently struggles in this dimension, as most works assume accurate CSI snapshots rather than imperfect or delayed channel information. Unlike mobility which can often be tracked through geometric models, channel variability presents deeper challenges. Addressing it calls for hybrid AI frameworks that combine online reinforcement learning with reliable channel-state estimation and proactive digital-twin simulation.

\subsection{Traffic load variability}
6G network traffic is likely to be very variable and service-diversified with its flow determined by coexisting eMBB, URLLC and mMTC requirements under industrial, vehicular and urban infrastructures. This variability imposes non-stationary traffic loads that put classical energy management under pressure due to typically resulting energy waste during off-peak periods or congestion during surges. AI-driven energy efficiency approaches have consequently to involve techniques that can predict traffic dynamics, vary resource utilization in real time and balance energy consumption and performance degradation.

Supervised learning and prediction-based methods have excellent flexibility to this dynamics at a proactive timescale. In industrial and smart city applications, CNNs, BPNNs and decision trees~\cite{paper87,paper92} are applied to learn load profiles and energy consumption tendencies allowing schedulers to activate or mute devices ahead of time. Digital twin technologies~\cite{dt3,dt4} take this a step further by modeling traffic variations before deployment allowing for long-term planning foresight and control of base station energy. Proactive techniques are efficient in highly structured environments, but are not flexible enough to respond to sudden spikes or bursty traffic.

Real-time adaptability is achieved through reinforcement learning agents that observe instantaneous traffic states such as queue lengths, congestion indicators or buffer status. RL-based solutions~\cite{paper98,paper47,paper90a} dynamically adjust base station activation, transmission power and load balancing minimizing energy waste while controlling throughput. In vehicular networks, duplex DRL and meta-reinforcement learning~\cite{v2,v4} further improve responsiveness by adapting to asymmetric or bursty loads. On the other hand, MARL~\cite{v3} distributes decision-making across vehicles and RSUs to sustain scalability under dense traffic.

However, solutions that are mostly based on physical-layer modifications, for example: UAV relocation~\cite{paper36,paper100} or RIS phase tuning~\cite{RIS1}, are less affected by long-term traffic fluctuations unless complemented by higher-layer schedulers. These methods implicitly adapt to load only through spatial reconfiguration, rather than explicit traffic-state inputs. 

In short, AI adaptability to traffic load encompasses both proactive and real-time approaches. Lightweight supervised predictors like CNNs, BPNNs and decision trees allow proactive scheduling in structured IoT and smart-city settings but their adaptability is limited to anticipated traffic patterns. Digital twin technologies extend this prediction to planning at the system level but are unable to reflect burstiness in real time. Stronger responsiveness arises from RL and MARL agents that adjust persistently to congestion and buffer states. In contrast, works centering on RIS- and UAVs almost entirely neglect temporal load dynamics characterizing then a missing piece in their design. By contrast to variability in mobility or channel, which evolve spatially or physically, traffic variability imposes a time-based volatility that requires a combination of proactive prediction with real-time scheduling via RL for efficient robust energy operation.

\subsection{QoS-driven delay and reliability requirements}

Energy-efficient 6G networks will not only reduce power consumption but also maintain stringent QoS levels with ultra-low latency, high reliability and minimum outage rates primarily in mission-critical applications such as V2X safety messaging, distant control in industrial IoT or instant UAV coverage. This puts a constant tension between energy-reduction choices and service-level guaranties. AI techniques will then have to learn how to strike a balance between utility maximization and compliance with QoS thresholds under dynamic traffic, channel and mobility states.

At the real-time timescale, reinforcement learning approaches have exhibited robust adaptability when their reward functions or constraints embeds QoS-awareness. The control of mode selection and power control based on DRL~\cite{paper16,v3,v4,v6} will dynamically adjust policies to reduce energy consumption without violation of latency or reliability limits. Actor–critic and multi-agent structures~\cite{paper37,paper47} also permit distributed agents to strike a balance between energy and QoS in dense or heterogeneous networks with a guarantee in fairness and stability. Nevertheless, these approaches typically depend on delicate reward shaping  and a wrong design can lead to energy-biased policies that degrade QoS. 

At a proactive timescale, supervised classification and scheduling models~\cite{paper87,paper90a} integrate service latency and criticality into their decision logic. In practice, these models give priority to urgent data flows while postponing less critical traffic. They work well in structured IIoT and smart city settings where service demands are fairly predictable. However, they struggle to react when QoS needs change suddenly, for instance, during an unexpected surge of emergency URLLC messages triggered by an accident or system failure. Digital twin frameworks~\cite{dt3,dt4} extend this by simulating service-level constraints in advance. In fact, they enable energy-aware deployment and orchestration strategies that incorporate reliability budgets. However, these remain long-term planning tools rather than reactive mechanisms.

On the other hand, physical layer-only solutions such as RIS reconfiguration or UAV trajectory planning without QoS feedback~\cite{RIS1,RIS2,paper36} show poor adaptability to service level guarantees. These techniques reduce spectral or coverage inefficiency wheras they risk to violate strict QoS targets when energy minimization is pursued separately.

In summary, AI adaptability to QoS constraints is highest when energy optimization is defined under specific service guarantees. Multi-agent and real-time RL schemes enable online responsiveness to reliability and latency requirements. However, DT-based and supervised approaches allow proactive priorities at slower timescales. The Worst adaptability occurs with purely physical-layer solutions ignoring QoS inputs. This highlights the need for hybrid approaches in which energy-saving policies are continuously checked and balanced against QoS metrics. Unlike traffic variability, which requires balancing temporal bursts, QoS dynamics impose hard service thresholds making constraint-aware RL and meta-RL particularly essential for sustainable 6G operation.

\subsection{Topological changes}
Topological variations in 6G networks are caused by dynamic relocation of infrastructure assets (e.g., UAVs, RIS panels), varying active node densities and changing link configurations in ad hoc or multi-hop deployments. Connectivity graphs, coverage areas and interference regions are changed, directly influencing the energy required for communication and coordination. Energy efficiency solutions based on AI must therefore respond with variable methods to these structural variations especially in scenarios that include moving base stations, reconfigurable surfaces or user-driven clustering.

At the real-time timescale, reinforcement learning has been shown to be efficient in reacting to topological variations. UAV trajectory control with Q-learning, PPO or double-DQN~\cite{paper36,paper100,RIS1,RIS4} enables aerial nodes to recalibrate their altitude, trajectory or transmit power continuously in accordance with changing coverage gaps or user clusters. In vehicular scenarios, multi-agent DRL structures~\cite{v3} scale to sparse and dynamic graphs by modeling agent interaction with mean-field approaches, allowing channel selection and power allocation to be quickly adapted as the graph changes. Similarly, RIS-assisted systems~\cite{RIS2,RIS4} adaptively adjust reflection matrices to achieve efficient communication in multi-hop or UAV-mounted configurations.

At a proactive timescale, industrial and smart city systems exploit clustering algorithms and DT-based planning to anticipate and mitigate structural shifts. Adaptive clustering and orchestration methods~\cite{paper44,paper45,paper87} reposition services or reassign roles when node density or demand hotspots change. DT frameworks~\cite{dt3,dt4} further simulate alternative layouts and optimize deployment strategies virtually, reducing energy costs of trial-and-error adjustments.

At a proactive timescale, industrial and smart city systems exploit clustering algorithms and DT-based planning to anticipate and mitigate topological shifts. Adaptive clustering and orchestration methods~\cite{paper44,paper45,paper87}  ensure services repositioning and roles reassignment when node density or demand hotspots are modified. DT frameworks~\cite{dt3,dt4} simulate alternative layouts and optimize deployment strategies virtually, reducing energy costs of trial-and-error adjustments.

Retrained or static methods such as supervised scheduling without spatial awareness~\cite{paper92} or RIS policies with fixed positions~\cite{RIS1} have poor adaptability to dynamics in topologies. They are efficient in steady states of layout, but must be manually retrained or reconfigured at regular times to efficiently work under a network graph that changes. 

In short, real-time adaptability to topological changes is strongest in RL- and MARL-driven UAV, V2X, and RIS-based solutions that explicitly embed spatial awareness. Proactive adaptability is supported by clustering and DT frameworks that anticipate density variations or layout changes before deployment. The weakest responsiveness is in static AI approaches that consider fixed topologies. Whereas traffic load or QoS dynamics are mostly time-oriented, topological dynamics redefine the very network structure itself, necessitating spatially enlightened agents that can learn and re-learn efficient energy policies as the environment re-configures.

In summary, real-time adaptability to topological changes is strongest in RL- and MARL-driven UAV, V2X, and RIS-assisted solutions that explicitly embed spatial awareness. Proactive adaptability is supported by clustering and DT frameworks that anticipate density variations or layout changes before deployment. The weakest responsiveness arises in static solutions that assume fixed topologies, highlighting the need for hybrid approaches that combine fast-reactive RL with proactive DT simulation. Unlike traffic load or QoS dynamics which mainly vary over time, topology dynamics change the actual connectivity of the network. They require spatially aware agents that can learn—and relearn—optimal energy policies as the environment reconfigures.

\subsection{Heterogeneity of service demands}

6G networks must simultaneously support diverse service classes with vastly different performance and energy requirements such as URLLC, eMBB and mMTC. These modes differ in their tolerance to delay, reliability, throughput and reporting frequency. In an energy optimization perspective, this heterogeneity creates a dual challenge: minimizing power consumption while respecting differentiated QoS constraints and dynamically adapting resource management to service-specific priorities.

At the real-time timescale, multi-agent DRL and hybrid control systems have demonstrated strong adaptability by conditioning policies on service-oriented QoS profiles. In V2X and IoT scenarios, MARL frameworks~\cite{v3,v5,v6} differentiate between delay-sensitive URLLC traffic and throughput-driven eMBB flows. This differentiation allows adaptive mode selection and power control that maintains crucial reliability while saving energy on less demanding traffic. Similarly, meta-learning and transfer learning approaches~\cite{v4,paper17} allow rapid reconfiguration of policies when new service types emerge or when application profiles change, addressing non-stationary demand mixes in real time.

At a proactive timescale, network slicing and classification models integrate service heterogeneity into orchestration. AI-driven slice management~\cite{paper90a} balances energy efficiency with service isolation by scheduling, migrating or resizing slices according to their latency or throughput requirements. In smart city and IoT deployments, supervised classification models~\cite{paper87,paper92} identify high-priority versus background traffic enabling schedulers to defer non-critical transmissions to save energy. Digital twin frameworks~\cite{dt3} extend this by simulating mixed traffic demands in virtual environments allowing operators to test allocation strategies before deployment.

Static or service-agnostic methods, such as RIS phase tuning or UAV repositioning without QoS or service-type input~\cite{RIS1,paper36} exhibit weak adaptability to heterogeneous demands. They optimize physical-layer performance but cannot ensure differentiated treatment of traffic classes making them less effective for environments where diverse service profiles coexist.

In a nutshell, adaptability to service heterogeneity is strongest in real-time MARL, transfer learning and meta-learning schemes that integrate service awareness with their learning cycle enabling context-aware power and scheduling decisions across URLLC, eMBB and mMTC. Proactive adaptability stems from slice-aware orchestration and supervised classification that anticipate service-level priorities. However, such models struggle when entirely new service classes emerge outside their training set. This reveals the gap between pre-configured service priorities and real-time diversity in 6G traffics. Service-agnostic physical-layer schemes show the lowest level of adaptability. This suggests that the diversity of service demands calls for AI models that explicitly account for service differentiation. Unlike traffic variability, which mostly changes over time, service heterogeneity introduces functional diversity. It requires policies that can continually balance competing service demands while maintaining energy efficiency.

\subsection{Resource constraints and availability}

6G networks must operate under strict limitations in energy supply, computational capacity and spectrum availability particularly in battery-powered IoT devices, UAVs with propulsion costs and dense access networks with overlapping slices. Energy-efficient operation in such constrained environments requires AI systems that not only optimize performance but also respect these hard physical and computational limits. They must adapt policies when available resources fluctuate or become scarce.

At the real-time timescale, RL and multi-agent coordination have shown high adaptability to resource-constrained environments. In SWIPT and spectrum reuse scenarios, actor–critic frameworks~\cite{paper37} continuously adjust sub-band allocation, transmit power and splitting ratios to maximize energy harvesting while preserving fairness. Similarly, RL-based base station controllers~\cite{paper98,paper47} learn to dynamically track BS activity and power levels according to buffer states and energy availability. As a result, they directly reduce consumption without compromising stability. In V2X networks, distributed MARL with federated extensions~\cite{v6} enables devices to share knowledge across agents while adapting to partial CSI and heterogeneous power budgets in real time.

At a proactive timescale, supervised models and digital twin frameworks anticipate resource limitations before they occur. Specifically, decision trees and random forests~\cite{paper92} classify device states (active, idle, near-depletion) to enable predictive scheduling. On the other hand, DT-based platforms~\cite{paper45,dt1} simulate power budgets and coverage trade-offs across factory or urban deployments. Consequently, they guide proactive allocation of spectrum and computing resources. These methods are effective in structured and predictable environments but cannot capture sudden resource shortages, for example sudden UAV propulsion drain or unexpected spectrum contention, situations where real-time learning is still essential.

In summary, AI adaptability to resource constraints is strongest thanks to real-time RL and MARL frameworks that embed energy, spectrum or computation directly into their reward structures enabling continuous trade-offs between utility and scarcity. On the other side, proactive models contribute by forecasting device depletion or simulating deployment-level budgets, but they remain limited in handling sudden shortages.

\subsection*{Adaptability Observations and Conclusions}

The evaluation of AI-driven energy efficiency strategies across the seven dynamics determines both strengths and persistent gaps. Reinforcement learning and its multi-agent extensions consistently emerge as the most effective tools for real-time adaptability. They allow UAVs, vehicles and base stations to react to mobility, channel variability and traffic surges with minimal predefined modeling. Meta-learning further strengthens this reactivity by accelerating policy updates under non-stationary conditions. At the proactive timescale, digital twin frameworks and supervised prediction models provide valuable foresight particularly in IIoT and smart city deployments. However, these proactive approaches fall short when confronted with sudden changes, leaving real-time agents as the only robust option in highly volatile settings. Overall, these observations underscore that no paradigm is enough on their own: sustainable energy efficiency in 6G will require hybrid AI frameworks that combine proactive planning with real-time responsiveness. Table~\ref{tab:usecase_dynamics} consolidates these findings offering a comparative view of how different use cases adapt across the identified dynamics.

\begin{table*}[t]
\centering
\caption{Adaptability of AI-driven EE solutions across 6G dynamics and use cases. 
Symbols: \cmark = strong real-time adaptability ; \tmarksym = partial or proactive adaptability ; \xmark = weak or static adaptability (fixed assumptions, retraining only).}
\rowcolors{1}{green!12!white}{green!2!white}
\arrayrulecolor{black}        
\setlength{\arrayrulewidth}{1pt} 
\renewcommand{\arraystretch}{1.15}
\small
\begin{tabular}{|p{1.7cm}|p{1.7cm}|p{1.7cm}|p{1.7cm}|p{1.7cm}|p{1.7cm}|p{1.7cm}|p{1.7cm}|}
\hline
\rowcolor{green!25}
\textbf{Use case} & \textbf{Mobility} & \textbf{Channel variability} & \textbf{Traffic load variability} & \textbf{QoS constraints} & \textbf{Topological changes} & \textbf{Service heterogeneity} & \textbf{Resource constraints} \\
\hline

\textbf{RIS-assisted communication} & 
\tmarksym (DRL for UAV-mounted RIS~\cite{RIS4}) & 
\cmark (DRL for phase optimization~\cite{RIS1,RIS2}) & 
\xmark (load rarely modeled) & 
\xmark (QoS not integrated~\cite{RIS1,RIS2}) & 
\tmarksym (mobile RIS scenarios~\cite{RIS4}) & 
\xmark (no service-type awareness) & 
\xmark (resource costs often ignored) \\
\hline

\textbf{V2X and high mobility} & 
\cmark (MARL, meta-RL for mode/power control~\cite{v3,v4,v6}) & 
\cmark (duplex DRL, meta-RL for Doppler~\cite{v2,v4}) & 
\cmark (RL/meta-RL for bursty loads~\cite{v2,v4}) & 
\cmark (constraint-aware RL for URLLC~\cite{v6}) & 
\cmark (mean-field MARL for dense topologies~\cite{v3}) & 
\cmark (MARL with service awareness~\cite{v5,v6}) & 
\tmarksym (federated MARL under partial CSI~\cite{v6}) \\
\hline

\textbf{UAV-assisted communication} & 
\cmark (RL for trajectory and power control~\cite{paper36,paper100}) & 
\cmark (DRL for channel-aware flight control~\cite{RIS4}) & 
\xmark (traffic not modeled explicitly) & 
\tmarksym (QoS only indirectly via outage~\cite{paper100}) & 
\cmark (RL for altitude/trajectory changes~\cite{paper36}) & 
\xmark (no service-type integration) & 
\cmark (power-aware RL under propulsion limits~\cite{RIS4}) \\
\hline

\textbf{IIoT and smart city} & 
\tmarksym (mobility prediction in NIB~\cite{paper17}) & 
\tmarksym (DT for channel forecasting~\cite{dt3,dt4}) & 
\cmark (CNN/BPNN for predictive scheduling~\cite{paper87,paper92}) & 
\tmarksym (supervised schedulers for latency-aware flows~\cite{paper87}) & 
\tmarksym (clustering + DT for deployment shifts~\cite{paper45}) & 
\cmark (supervised classification of traffic~\cite{paper87,paper92}) & 
\cmark (DTs for energy budget simulation~\cite{dt1,dt3}) \\
\hline

\textbf{Network operation \& resource allocation} & 
\tmarksym (BS control agents with limited mobility input~\cite{paper47}) & 
\tmarksym (actor–critic using channel feedback~\cite{paper37}) & 
\cmark (RL for BS sleep/wake~\cite{paper98,paper90a}) & 
\cmark (actor–critic MARL for QoS-energy trade-offs~\cite{paper37}) & 
\tmarksym (adaptive BS switching under load~\cite{paper98}) & 
\cmark (slice-aware orchestration~\cite{paper90a}) & 
\cmark (RL for spectrum and power budgets~\cite{paper37}) \\
\hline

\end{tabular}
\label{tab:usecase_dynamics}
\end{table*}

\section{Core tradeoffs and AI strategies in energy-aware 6G Design}

Whereas AI methods present powerful capabilities to minimize energy consumption in 6G networks, their implementation frequently introduces tensions among performance, complexity and system-level considerations. The tradeoffs arise as a consequence of structural limitations (e.g., constrained communication or computational capacity), application-aware constraints (e.g., spatial coverage or real-time inference) along with the dynamic nature of the 6G environment. Knowing how AI strategies address these tradeoffs is crucial to the development of sustainable and dynamic energy-aware systems. This section characterizes six influential tradeoffs that are commonly faced across the spectrum of the energy-aware use cases in 6G and discusses how various AI methodologies are utilized to resolve them. For each of the tradeoffs, we examine the management strategies, the AI technique role as well as the use case scenarios wherein the tension is most significant.

\subsection{Energy efficiency vs. Network performance (QoS, Latency, Reliability)}
In 6G networks, ensuring energy efficiency while maintaining acceptable levels of performance particularly in terms of Quality of Service, latency, reliability and throughput poses a significant optimization challenge. The inherently dynamic and heterogeneous nature of 6G exacerbates this tradeoff necessitating then intelligent and context-aware mechanisms. AI techniques provide a powerful means to balance this tradeoff through adaptive learning, multi-objective optimization, and real-time decision making. Based on the reviewed literature, several distinct approaches emerge for how AI manages this balancing.

A prominent strategy involves the use of reinforcement learning frameworks, where the reward function is explicitly designed to balance energy consumption and QoS compliance. For instance, in UAV-assisted networks~\cite{paper36}, researchers incorporate penalty terms into the reward function that discourage violations of minimum data rate thresholds. The RL agent thus learns to avoid low-quality coverage zones while optimizing UAV energy consumption. Some works further extend this by incorporating adjustable penalty weights or dynamic reward scaling that allows systems to change emphasis between performance and energy objectives based on evolving conditions~\cite{paper100,RIS3,RIS4}.

Another approach relies on multi-objective optimization techniques that treat energy efficiency and network performance as separate objectives to be jointly optimized. These formulations often employ scalarization techniques such as weighted sum or Tchebycheff methods \cite{paper37} which aims to minimize the maximum deviation from each ideal objective (e.g., energy and latency). This allows the AI agent to evaluate not just the sum of objectives but their worst-case satisfaction. Thus, it ensures a more robust balancing under fluctuating network conditions. This method is particularly useful in vehicular networks and ultra-reliable low-latency communication scenarios.

Parameterized control schemes also offer flexible mechanisms to manage this tradeoff. In such models, a single tunable parameter controls the operational mode of the system allowing, for example, more aggressive energy saving during off-peak hours and tighter QoS adherence during high-load periods. For example, in base station sleep mode management~\cite{paper98}, a parameter determines the balance between entering deep sleep states (to save energy) versus maintaining low latency (to preserve QoS). By adjusting this parameter, the system can switch modes of operation from energy-focused to performance-focused based on contextual needs.

In learning-based systems especially in supervised or deep learning architectures, the tradeoff is often handled implicitly. Models are trained on datasets that embed historical tradeoff decisions or are fine-tuned using validation strategies that include both energy and performance metrics. For instance, in hybrid deep learning models applied to dynamic network slicing~\cite{paper90a}, CNNs and RNNs jointly learn to allocate resources in a way that reduces energy usage while ensuring throughput and latency targets are not compromised.

In multi-agent settings, the tradeoff is managed through decentralized coordination where each agent balances local energy consumption with its contribution to global service quality. Agents share environmental observations or partial policy states to align their behaviors in order to achieve network-wide efficiency while maintaining distributed performance guarantees~\cite{RIS2,paper81a}. These systems are particularly effective in massive IoT environments, where centralized optimization is infeasible and local decisions must be adjusted with global efficiency goals.
Finally, in some advanced frameworks, digital twin systems combined with DRL are used to simulate and evaluate the impact of different strategies on both energy and QoS before deployment~\cite{paper45}. These predictive models enable proactive tradeoff management by anticipating when energy-saving actions may degrade service quality and vice versa.

In summary, AI handles the energy efficiency vs. performance trade-off through a wide range of mechanisms ranging from explicit multiobjective reward design and parameter tuning to implicit learning from data and distributed coordination. The choice of approach depends on the application domain, the dynamics of the network and the required level of adaptability.

\subsection{Energy efficiency vs. Computational complexity}

Achieving energy efficiency in 6G networks often requires solving complex, high-dimensional optimization problems. This optimization involves multiple parameters including user association, resource scheduling, beamforming or mobility control. However, optimizing energy consumption through AI methods introduces a tradeoff with computational complexity both during training and inference phases. This tradeoff becomes particularly significant in real-time, distributed or resource-constrained environments such as industrial IoT, UAV swarms and RIS-assisted systems. AI techniques thus face the dual challenge of minimizing energy at the network level while remaining computationally sustainable at the algorithmic level. 

A widely adopted strategy involves the use of lightweight reinforcement learning methods or simplified state-action spaces. Instead of relying on high-dimensional representations, models abstract system dynamics into compressed state vectors or reduced action sets to limit training overhead. For example, tabular Q-learning variants, SARSA models and state aggregation techniques are used in scenarios like base station activity control and power allocation. These effectively reduce convergence time and memory demands~\cite{paper98}.

Another interesting solution leverages neural network quantization and pruning to reduce the computational burden of inference. In edge-deployed AI systems particularly for BS energy control or RIS configuration, quantized DNNs offer substantial reductions in processing time and power consumption without significantly sacrificing decision accuracy. This makes them particularly suitable for massive IoT deployments, where inference must be fast and scalable~\cite{paper81a}.

Some frameworks utilize a hierarchical learning design that separates global high-level policies from localized low-complexity agents. This modular structure allows computationally intensive components (e.g., trajectory planning, global coordination) to run in centralized cloud controllers while lightweight agents handle localized real-time adaptation. This layered approach reduces edge-side complexity while preserving global energy optimization~\cite{RIS2}.

In other scenarios, deep learning models are combined with rule-based or threshold-triggered mechanisms in order to offload complexity. The systems will use the learned models at the situation of uncertainty or dynamic changes instead of calling full AI-led decisions at each time step. This kind of combination is beneficial in latency-constrained applications like UAV relaying or mobile MEC where the delay of decision should be kept as minimum as possible~\cite{RIS4, paper87}.

A few approaches exploit transfer learning or meta-learning to pre-train energy-efficient policies offline and deploy them quickly in new environments with minimal retraining. These strategies significantly reduce online computation and enable fast adaptation to topological or traffic variations while still maintaining energy-aware behavior~\cite{v4}.

In summary, AI methods address the tradeoff between energy efficiency and computational complexity through model simplification, hierarchical learning, quantization and hybrid decision mechanisms. The choice of strategy depends on the deployment context: lightweight designs are essential for edge-based control while centralized architectures allow richer optimization at the cost of delay. Balancing algorithmic efficiency with system-level energy goals represents a key design principle for AI in 6G.

\subsection{Energy Efficiency vs. User fairness and Resource equity}
Optimizing goals in energy-efficient 6G systems typically are to reduce overall power consumption or to improve global energy utility. Such global performance measures can, however, inadvertently cause resource unbalancing, such that edge users, low-mobility devices and nodes with poor channel are deprived of performance or even service starvation. This poses a significant tradeoff between the optimization of the overall system efficiency as well as fair energy consumption and service provision to users and links. The high heterogeneity and density of 6G environments spanning IoT, UAVs and RIS-assisted architectures make this tradeoff especially particular.

One common strategy includes fairness constraints directly into the learning or optimization objective. Reinforcement learning agents are trained using reward functions involving fairness indicators such as proportional throughput, energy balance or Jain's fairness index \cite{jain} to prevent policy convergence toward resource-hungry or dominant users~\cite{paper47, paper87}. This helps ensure that energy-efficient decisions also preserve minimum service guarantees for disadvantaged nodes. Another approach leverages MARL where each user or device is treated as an autonomous agent. By enabling cooperative policies or role-aware learning, the system distributes energy-saving responsibilities more equitably allowing agents to negotiate tradeoffs locally while aligning with global efficiency goals~\cite{RIS2, paper81a}. This is particularly effective in decentralized IoT or RIS-enabled deployments where centralized control is not practical.

Scheduling-based strategies also contribute to fairness by prioritizing under-served users during periods of resource surplus. AI models learn context-aware scheduling rules that ensure periodic access to energy-saving opportunities for all participants. These rules may be statically defined or learned through imitation or meta-learning techniques, allowing flexible re-balancing under changing load patterns~\cite{paper47, RIS1}.

Other solutions use hybrid schemes with user classification (e.g., elastic vs. critical traffic) dictating the form of the energy-aware action. For instance, in BS clustering or task offloading, elastic users may be subjected to more aggressive energy reduction policies while critical users receive preferential access to energy-hungry but latency-optimal pathways. This class of strategies balances fairness with application-aware performance guarantees~\cite{paper87,paper100}.

In summary, addressing the energy efficiency vs. fairness tradeoff requires embedding equity into the AI control loop either through explicit objective functions, decentralized collaboration or adaptive user-aware policies. As 6G systems evolve toward ultra-dense, service-diverse deployments, ensuring that energy-saving strategies do not come at the expense of under-served or peripheral nodes becomes a critical requirement for sustainable and inclusive network intelligence.

\subsection{Energy efficiency vs. Task utility and inference accuracy}

In many 6G scenarios, particularly those involving sensing, decision making or distributed AI, energy-saving measures can directly impact the utility of the data collected or the accuracy of inference tasks. For instance, disabling sensors, limiting sampling rates, pruning features or reducing computation cycles may reduce energy usage but also compromise the quality or the fidelity of the information used for communication, control or prediction. This introduces a key tradeoff: optimizing energy efficiency must not significantly degrade the effectiveness of the tasks that the system is meant to support. As 6G moves toward ubiquitous intelligence, this tradeoff becomes increasingly important especially in autonomous vehicles, smart factories and edge inference pipelines.

A prominent strategy for managing this tradeoff is strategically selecting the best features or prioritizing some sensors. In other words, AI models are trained to identify and retain only the most informative features or sensor streams. As a result, energy-intensive data acquisition and transmission are reduced while preserving decision quality. These methods are often implemented via reinforcement learning or supervised pruning techniques, where task performance metrics such as classification accuracy or latency are monitored to guide selection~\cite{paper45, paper92}.

Another approach relies on model compression and approximation techniques. By replacing complex inference models with lightweight surrogates such as quantized neural networks, shallow classifiers or regression-based estimators, systems achieve real-time performance under energy constraints. These approximate models are often fine-tuned to retain critical task-specific accuracy while reducing computation and memory usage~\cite{paper81a}.

Federated learning schemes also manage this tradeoff by enabling partial model updates or selective aggregation. Rather than performing full high-resolution inference at each node, devices may offload only critical gradient information or low-dimensional representations. This allows devices to reduce local computation and communication overhead. When combined with energy-aware client selection or participation thresholds, such systems balance utility with cost~\cite{v6, paper49}.

In addition, predictive control or early exit mechanisms can be deployed. In fact, deep models are equipped with intermediate classifiers or confidence thresholds that allow them to complete inference early when sufficient accuracy is achieved, thus saving computation cycles. This dynamic inference strategy aligns energy usage with task certainty and is especially useful in event-driven systems or latency-constrained applications~\cite{RIS4}.

In summary, AI-enabled systems handle the tradeoff between energy efficiency and task utility through selective data usage, model simplification, early stopping and collaborative processing. The most effective solutions adaptively match resource allocation to task difficulty and importance. It ensures that energy reductions do not come at the cost of critical functional accuracy. This is crucial for maintaining trust and reliability in energy-constrained 6G intelligence.

\subsection{Energy Efficiency vs. Spatial Coverage and Continuity}

In 6G networks, spatial coverage and seamless connectivity are fundamental to support mobile users, dense urban environments and emerging use cases. However, actions taken to improve energy efficiency such as deactivating base stations, limiting UAV movement or constraining RIS configurations can lead to coverage holes, increased handover frequency or degraded signal quality. This introduces a fundamental tradeoff: reducing energy expenditure must not achieve spatial service continuity particularly in scenarios that requires ubiquitous and uninterrupted communication. 

A key strategy involves mobility-aware reinforcement learning where UAVs, ground nodes and RIS elements are dynamically positioned to optimize both energy use and coverage quality. The RL agents are trained using reward functions that penalize coverage gaps, poor connections or excessive handovers. This strategy guaranties that energy-saving trajectories or configurations maintain sufficient available area-wide service~\cite{paper100,RIS1}.

Multi-objective optimization also plays a role especially in scenarios with large-scale spatial domains. Practically, these formulations explicitly model both energy consumption and spatial coverage metrics (e.g., SINR maps, coverage probability) as competing objectives. AI models learn policies that optimize over Pareto frontiers allowing that energy-efficient configurations do not excessively shrink service areas~\cite{RIS2, paper90a}.

In RIS-assisted systems, coverage-aware phase shift control is employed to dynamically reconfigure the reflection environment. AI algorithms determine optimal RIS settings not only to minimize transmit power but also to preserve line-of-sight or non-line-of-sight coverage paths even as users move or environments change. This maintains consistent signal reflection while avoiding unnecessary phase element activations~\cite{RIS3}.

Hierarchic or hybrid architectures also assist in this tradeoff management by distinguishing long-term coverage planning from short-term control of energy. A high-level planner maintains spatiotemporal continuity and a low-level controller insists on energy savings within that direction. This distinction enables anticipatory control that maintains user experience while cultivating efficiency~\cite{RIS4}.

Overall, the tradeoff between spatial coverage and energy efficiency necessitates that AI systems incorporate location awareness, mobility patterns of users and environmental variability within their decision. Either with the help of mobile node control based on adaptation, intelligent surface configuration or predictive coverage simulation, effective solutions guarantee that energy-saving actions do not compromise the reliability and continuity of 6G service footprints.

\subsection{Tradeoff: Energy Efficiency vs. Adaptability and learning overhead}

Achieving energy efficiency in 6G networks often demands context-aware and adaptive AI models capable of reacting to dynamic environments, mobility, and user heterogeneity. However, adaptability comes at a cost: learning processes especially in online, federated or multi-agent settings—incur communication, computation and storage overhead. These overheads may offset the energy gains they seek to enable, particularly in distributed, resource-constrained systems such as IoT, UAVs and mobile edge networks. This gives rise to a fundamental tradeoff: balancing the energy costs of learning and adaptation against their long-term benefits in dynamic optimization.

One prominent strategy involves leveraging meta-learning techniques~\cite{meta1, meta2, meta3} to reduce the cost of repeated training. Instead of learning from scratch in each new environment, AI agents are trained to learn quickly with minimal data or compute resources. Meta-RL, in particular, has shown promise in enabling agents to adapt to shifting QoS or mobility conditions while minimizing online training costs~\cite{v4}. These methods improve adaptability without extensive retraining, making them suitable for fast-changing 6G contexts.

In distributed learning environments, techniques such as selective model updates, asynchronous training and client sampling are employed to limit communication rounds and computational load~\cite{v6,paper49}. These strategies are especially relevant to federated learning scenarios where energy-constrained devices must contribute to global optimization without incurring excessive participation costs.

In addition, there are some systems that use periodic learning with dynamic scheduling: models are only updated when system drift is detected, or when performance degrades beyond a threshold. This avoids unnecessary retraining in stable periods reducing both energy and latency overhead.

In summary, managing the tradeoff between energy efficiency and adaptability overhead requires a careful orchestration of when, where and how learning occurs. AI solutions must optimize not only for performance and power but also for the efficiency of their own adaptation processes. Future 6G deployments will benefit most from systems that treat adaptability itself as a constrained resource to be optimized.

Overall, the tradeoffs that we have explored demonstrate that adaptability in AI-enabled energy efficiency solutions cannot be separated from the balancing of competing system goals.  Many of the same strategies, reward shaping in the context of reinforcement learning, multi-objective optimization, hierarchical decision architectures and hybrid control reappear over varied tradeoffs as well as varied deployment scenarios. This repetition mirrors the versatility of these AI methodologies to navigate the shifting 6G landscape where structural constraints, service heterogeneity and topological dynamics constantly interact. Collectively, the tradeoffs as well as the adaptability assessment reinforce that 6G sustainability in the form of energy-aware intelligence in 6G is not about optimizing a single metric in isolation. Rather, it is about managing tensions across objectives while remaining responsive to dynamic network conditions.

\begin{table*}[ht]
\centering
\captionsetup{font=small,aboveskip=2pt,belowskip=4pt}
\rowcolors{1}{green!12!white}{green!2!white}
\arrayrulecolor{black}        
\setlength{\arrayrulewidth}{1pt}  
\caption{Core tradeoffs in energy-efficient 6G and AI strategies to address them}
\begin{tabular}{|p{3.5cm}|p{4.2cm}|p{4.2cm}|p{3.5cm}|}
\hline
\textbf{Tradeoff} & \textbf{AI Techniques / Strategies} & \textbf{Role of AI Technique} & \textbf{Key 6G Use Cases} \\
\hline
\textbf{EE vs. Network Performance} & QoS-aware RL, Multi-objective optimization, CNN+RNN hybrid models & Balance energy saving with throughput, latency, and reliability under dynamic constraints & UAV coverage, network slicing, BS mode control \\
\hline
\textbf{EE vs. User Fairness and Resource Equity} & Fairness-aware RL, Multi-agent coordination, Role-aware scheduling & Ensure equitable service and avoid energy starvation for weak or edge users & BS clustering, D2D in RIS, IoT device management \\
\hline
\textbf{EE vs. Task Utility and Inference Accuracy} & Feature pruning, Quantized DNNs, Confidence-based early exits & Preserve learning or sensing performance while reducing sensing or computation energy & Edge inference, smart sensing, collaborative analytics \\
\hline
\textbf{EE vs. Spatial Coverage and Continuity} & Mobility-aware RL, Hierarchical UAV control, Coverage-aware RIS optimization & Maintain signal availability and user continuity while minimizing movement and active resources & UAV relay networks, RIS-assisted MEC, wide-area IoT \\
\hline
\textbf{EE vs. Adaptability and Learning Overhead} & Meta-learning, Lightweight RL agents, Federated model update strategies & Enable real-time adaptation without excessive training or communication costs & Distributed MEC, federated IoT, context-aware edge AI \\
\hline
\textbf{EE vs. Computational Complexity} & Simplified RL, Quantized models, Rule-based hybrid decision logic & Reduce inference and training burden to enable deployment in constrained settings & BS control at the edge, RIS reconfiguration, dense device coordination \\
\hline
\end{tabular}
\label{tab:ee_tradeoffs_summary}
\end{table*}

\begin{table*}[ht]
\centering
\captionsetup{font=small,aboveskip=2pt,belowskip=4pt}
\rowcolors{1}{green!12!white}{green!2!white}
\arrayrulecolor{black}        
\setlength{\arrayrulewidth}{1pt} 
\renewcommand{\arraystretch}{1.15}
\small
\caption{AI techniques and the tradeoffs they manage in energy-efficient 6G networks}
\begin{tabular}{|p{3.2cm}|p{4.3cm}|p{3.4cm}|p{4.2cm}|}
\hline
\textbf{AI Technique / Strategy} & \textbf{Typical Tradeoffs Managed} & \textbf{Core EE Benefit} & \textbf{Limitations / Tradeoff Sensitivity} \\
\hline
\textbf{QoS-aware Reinforcement Learning} & EE vs. Network Performance, EE vs. Spatial Coverage & Dynamically adjusts actions to meet QoS under energy constraints & Sensitive to reward design; slow adaptation to rapid dynamics \\
\hline
\textbf{Multi-objective Optimization (RL/DL)} & EE vs. Network Performance, EE vs. Coverage & Balances conflicting goals (e.g., energy vs. latency or coverage) via Pareto-aware policy learning & High computational cost in large-scale or real-time systems \\
\hline
\textbf{Fairness-aware Reward Shaping (RL)} & EE vs. User Fairness & Encourages energy efficiency without starving edge or weak users & Requires balancing system utility vs. individual fairness \\
\hline
\textbf{Multi-agent Reinforcement Learning (MARL)} & EE vs. User Fairness, EE vs. Learning Overhead & Enables distributed, scalable EE decisions across heterogeneous nodes & Coordination overhead and partial observability challenges \\
\hline
\textbf{CNN + RNN Architectures (Hybrid DL)} & EE vs. Network Performance & Learns spatio-temporal energy control patterns without explicit programming & Needs high-quality training data; inference cost can grow with model depth \\
\hline
\textbf{Model Compression / Quantization} & EE vs. Computational Complexity, EE vs. Task Utility & Reduces inference and deployment energy for on-device or edge models & Risk of degrading accuracy or model stability \\
\hline
\textbf{Feature Pruning / Sensor Selection} & EE vs. Task Utility & Drops low-utility data sources to reduce sensing and transmission power & Task performance may degrade if features are wrongly discarded \\
\hline
\textbf{Meta-learning / Model Reuse} & EE vs. Learning Overhead, EE vs. Fairness & Speeds up adaptation across changing environments with reduced retraining cost & Requires prior knowledge of task distribution and training diversity \\
\hline
\textbf{Hierarchical Control / Two-level Agents} & EE vs. Learning Overhead, EE vs. Spatial Coverage & Splits complex EE decisions into fast local actions and strategic global planning & Coordination logic can be complex and hard to generalize \\
\hline
\textbf{Early-exit~/ Confidence-aware Inference} & EE vs. Task Utility & Avoids full model inference when high-confidence predictions are achieved early & Requires dynamic calibration; risk of premature exit errors \\
\hline
\end{tabular}
\label{tab:ai_tradeoffs}

\end{table*}
\subsection{Key Observations and Lessons Learned}

Based on both the adaptability study and the tradeoff assessment, several recurring patterns and design lessons emerge in AI-based energy optimization for 6G. First, reinforcement learning and multi-agent methods consistently come out as the most adaptable strategies. They effectively handle tradeoffs involving QoS, fairness and spatial coverage across diverse deployment settings. Although effective in structured environments or static tasks, supervised and hybrid deep learning approaches often lack the flexibility needed to adapt to dynamic network conditions. Across all categories, the most robust solutions are those that explicitly encode tradeoff structures into their objectives either through multi-objective formulations, reward shaping or role-aware coordination rather than optimizing energy separately. Additionally, the rise of lightweight inference models, hierarchical controllers and meta-learning strategies points toward a growing recognition of the computational and adaptability burdens imposed by learning itself. The studies point to a crucial transformation: AI is no longer a device to minimize energy but a decisional level that must balance energy with performance, fairness and utility with the dynamic 6G evolution. Designing for this balance, rather than for static energy goals, represents a key lesson for sustainable and intelligent 6G architectures.

\section{Gaps and Future Directions}

Despite substantial progress in using AI to optimize energy efficiency in 6G networks, several research gaps remain that limit robustness of the scalability and generalizability of the current solutions. The reviewed literature reveals promising trends such as reinforcement learning for adaptive control, deep learning for spatiotemporal optimization and multi-agent systems for decentralized coordination, but also exposes critical limitations and underexplored dimensions.

\textbf{1. Limited generalization across scenarios.}  
Many proposed AI models are tightly coupled to specific environments, topologies or traffic patterns. This constrains their applicability in highly dynamic and heterogeneous 6G ecosystems. Generalization across mobility levels, application types, and user densities remains a key challenge particularly for supervised and deep learning models~\cite{paper36, paper90a}. Future research should explore more robust meta-learning, domain adaptation and simulation-to-reality transfer to build reusable and context-flexible AI agents.

\textbf{2. Tradeoff oblivious optimization objectives.} 
Much of the existing work targets optimization of energy consumption implicitly or with assumption of easy QoS or fairness constraints. Fewer models actually tackle multi-objective or tradeoff-aware optimization explicitly~\cite{paper47,paper90a,RIS2}. The building of AI architectures that natively cope with tradeoff surfaces through scalarization, hierarchical control or adaptation of priorities remains a valid direction. Of no smaller importance is the establishment of standardized benchmarks and KPIs to be used to assess the trade-off between energy-QoS-fairness in use cases.

A significant portion of the current work focuses on optimizing energy consumption alone or assumes simplified QoS or fairness constraints. Only a minority of models explicitly address multi-objective or tradeoff-aware optimization~\cite{paper47,paper90a,RIS2}. Developing AI architectures that are natively designed to handle tradeoff surfaces via scalarization, hierarchical control or adaptive prioritization remains an open direction. Equally important is the development of standardized benchmarks and KPIs to evaluate the trade-offs between energy efficiency, quality of service and fairness across different use cases.

\textbf{3. Underexplored resource-constrained learning.}  
AI-based energy optimization methods often incur notable overhead during training or inference limiting then their deployment in real-time or low-power environments like massive IoT or edge devices~\cite{paper81a,paper87}. Lightweight AI design through model compression, continual learning and efficient search strategies deserves deeper attention to improve deployability and responsiveness. This gap extends to the energy footprint of AI itself, where pruning, quantization and distillation must be co-optimized with network-level EE targets.

\textbf{4. Lack of joint learning across layers and domains.}  
Most reviewed strategies operate at a single layer (e.g., PHY, MAC, or application) or focus on isolated decisions such as BS sleep control or task offloading. However, 6G efficiency is inherently cross-layer and multi-domain. There is a pressing need for integrated learning frameworks that coordinate decisions across networking, computing and sensing domains, especially in scenarios involving RIS, UAVs and MEC~\cite{RIS1,RIS4,paper100}.

\textbf{5. Scarcity of realistic evaluation.} Most studies rely on simulations with idealized assumptions (e.g., perfect CSI, fixed user models). Furthermore, little effort is made to interpret how AI decisions align with energy–performance tradeoffs. Future directions include incorporating realistic mobility, uncertainty and hardware constraints alongside explainable AI (XAI) to make energy-aware policies transparent and trustworthy.

\textbf{6. Dynamic tradeoff management remains immature.}  
Although some works introduce tunable parameters or context-aware reward shaping, few systems dynamically adapt their optimization emphasis based on external stimuli (e.g., emergency load, user priority or mission phase). Research into self-adjusting tradeoff frameworks perhaps guided by reinforcement meta-control or user feedback could offer more elastic and resilient energy strategies.

\textbf{7. Hybrid adaptation across timescales.}  
Current solutions are polarized between reactive real-time control (e.g., RL agents) and proactive forecasting (e.g., supervised prediction or DT simulation). Few frameworks combine the two. Future research should explore hybrid adaptation pipelines where proactive DT foresight is fused with real-time RL policies allowing networks to anticipate demand trends yet also react instantly to sudden changes.

\textbf{8. Resilience under partial observability.}  
Many AI solutions implicitly assume perfect CSI, complete traffic knowledge or accurate environment sensing. In practice, sensing may be delayed, incomplete or noisy, especially in UAV, RIS and vehicular deployments. Future work should explore Bayesian RL, distributionally robust optimization and safe exploration techniques to ensure that AI-driven energy control remains effective under imperfect observability.

\vspace{0.5em}
In summary, while AI has unlocked new possibilities for dynamic, context-aware energy optimization in 6G, its current application remains fragmented and often domain-specific. The path forward lies in unifying AI robustness, tradeoff intelligence, hybrid adaptation across timescales and resilience to uncertainty paving the way for scalable, transparent and energy-aware intelligence at the heart of next-generation networks.

\section*{Conclusion}
The pursuit of energy efficiency in 6G networks is inseparable from the need to operate in dynamic, heterogeneous and unpredictable environments. This survey has reviewed a broad range of AI-driven approaches designed to optimize energy use while sustaining the performance requirements of diverse 6G use cases including RIS-assisted communications, industrial IoT and smart cities, UAV networks, V2X, and large-scale resource management. Across these use cases, reinforcement learning, multi-agent collaboration, predictive modeling and digital twins have consistently emerged as enablers of adaptability allowing networks to sense environmental changes and autonomously reconfigure resources.

Our adaptability evaluation shows that no single technique provides universal resilience across all dynamics. Real-time reinforcement learning excels in environments dominated by fast-changing mobility and channel conditions. Proactive digital twin and prediction-based methods offer value in longer-term orchestration and deployment planning. However, the interplay of these approaches remains underdeveloped. It highlights the importance of hybrid adaptation strategies. Furthermore, the analysis highlights open challenges in generalization, tradeoff intelligence, lightweight learning and robustness under imperfect observability.

Looking forward, the integration of AI with sustainability principles will require cross-layer and cross-domain frameworks capable of managing tradeoffs dynamically and transparently. Edge-enabled intelligence, federated and distributed learning, and explainable decision-making will be critical to ensure that AI not only optimizes local energy usage but also contributes to the global carbon-neutral objectives of 6G. Interestingly, adaptability must be embedded as a foundational property of AI-based solutions guaranteeing that future wireless networks achieve energy efficiency without compromising reliability, scalability or fairness. This vision requires the unification of adaptive AI design with systemic 6G architecture, positioning energy-aware intelligence at the core of next-generation networks.

\bibliographystyle{cas-model2-names}

\bibliography{biblio}

\section*{List of Acronyms}

\renewcommand{\arraystretch}{1.15}
\setlength{\tabcolsep}{6pt}
\begin{tabularx}{\columnwidth}{lX}
\hline
\textbf{Acronym} & \textbf{Definition} \\
\hline
6G    & Sixth Generation Mobile Network \\
AI    & Artificial Intelligence \\
BPNN  & Backpropagation Neural Network \\
CNN   & Convolutional Neural Network \\
D2D   & Device-to-Device \\
DDPG  & Deep Deterministic Policy Gradient \\
DDQN  & Double Deep Q-Network \\
DQN   & Deep Q-Network \\
Double-DQN & Double Deep Q-Network \\
DRL   & Deep Reinforcement Learning \\
DT    & Digital Twin \\
EE    & Energy Efficiency \\
EDDL  & Environment-Driven Deep Learning \\
eMBB  & Enhanced Mobile Broadband \\
FL    & Federated Learning \\
IIoT  & Industrial Internet of Things \\
ITS   & Intelligent Transportation Systems \\
LSTM  & Long Short-Term Memory \\
MARL  & Multi-Agent Reinforcement Learning \\
ML    & Machine Learning \\
MEC   & Multi-access Edge Computing \\
mMTC  & Massive Machine Type Communications \\
NIB   & Network-in-Box \\
NOMA  & Non-Orthogonal Multiple Access \\
PPO   & Proximal Policy Optimization \\
QoE   & Quality of Experience \\
QoS   & Quality of Service \\
RIS   & Reconfigurable Intelligent Surface \\
RL    & Reinforcement Learning \\
RNN   & Recurrent Neural Network \\
SARSA & State–Action–Reward–State–Action \\
UAV   & Unmanned Aerial Vehicle \\
URLLC & Ultra-Reliable Low-Latency Communications \\
V2X   & Vehicle-to-Everything \\
\hline
\end{tabularx}


\end{document}